\documentstyle[manuscript,epsfig,aps]{revtex}
\begin{document}
\draft

\title{Linking Phase-Field and Atomistic Simulations to
Model Dendritic Solidification in Highly Undercooled Melts} 

\author{Jean Bragard$^1$, Alain Karma$^1$,
Youngyih H. Lee$^{1,*}$, and Mathis Plapp$^2$}
\address{
$^1$Department of Physics and Center for Interdisciplinary
Research on Complex Systems,
Northeastern University,
Boston, MA 02115,
USA\\
$^2$Laboratoire de Physique de la Mati\`ere Condens\'ee,
CNRS/Ecole Polytechnique, 91128 Palaiseau, France
}

\date{\today}
\maketitle

\begin{abstract}
Even though our theoretical understanding of dendritic
solidification is relatively well developed, our current ability to model
this process quantitatively remains extremely limited. 
This is due to the fact that the morphological development of dendrites  
depends sensitively on the degree of anisotropy of
capillary and/or kinetic properties of the solid-liquid interface, 
which is not precisely known for materials of metallurgical interest. 
Here we simulate the crystallization of highly
undercooled nickel melts using a computationally efficient
phase-field model together with anisotropic properties
recently predicted by molecular dynamics simulations. 
The results are compared to experimental data and to the predictions of 
a linearized solvability theory that includes both  
capillary and kinetic effects at the interface.
\end{abstract}

\vskip 1 cm
\noindent {\small $^*$Present address: Department of Radiation Oncology,
Yonsei University, College of Medicine,
Seoul 132, Korea.}

\pagebreak

\section{Introduction and overview}

The freezing of a supercooled liquid or a supersaturated
liquid mixture of two or more components generally leads to
the formation of highly branched dendritic crystals in  
materials where the solid-liquid interface is rough on an
atomic scale, but smooth on a macroscopic scale \cite{TriKar}. 
This process
is of considerable practical importance to metallurgists because
commercial metallic alloys often form dendrites during freezing, and
the resulting microstructure controls the properties
of the final solidification product \cite{KurFis}.
Moreover, dendritic growth
has been of fundamental interest to physicists in the general context
of pattern formation in non-equilibrium dissipative systems 
\cite{Kesetal,Lan:Houches,Kar:Houches}. 

The solidification of a pure undercooled melt is one of
the simplest methods to form dendritic crystals and to investigate their
growth behavior and morphology under well-controlled conditions.
For this reason, this process has been extensively 
studied both theoretically and experimentally 
for several decades. Traditionally, experimental studies
have focused on the characterization of the unique  
relationship between the steady-state growth velocity $V$ 
of the dendrite tip and the bulk undercooling 
$\Delta T=T_M-T_\infty$, where $T_M$ is the melting point temperature
and $T_\infty$ is the initial temperature of the melt.  

A first major step in the development 
of the theory of dendritic growth was made over half a century ago
by Ivantsov \cite{Iva} who analyzed this problem in the simplifying
limit where both capillary and kinetic 
effects at the interface are neglected. In this approximation, 
the solid-liquid interface grows  
at the melting temperature and the only
rate-limiting process is the diffusive transport of the latent heat of freezing 
into the undercooled liquid. The basic equations governing the
growth are then the diffusion equation
\begin{equation}
\frac{\partial T}{\partial t}=D\nabla^2 T,\label{e1}
\end{equation} 
and the condition of heat conservation at the 
growing interface
\begin{equation}
LV_n=c_pD\,\hat n\cdot \left(\left.\vec\nabla T\right|_S
-\left. \vec\nabla T\right|_L\right) \label{e2},
\end{equation} 
where $V_n$ is the normal velocity of the interface, $\hat n$
is the directional normal to the interface pointing into the liquid and
$\hat n\cdot \vec\nabla T|_{S,L}$ is the normal gradient
of temperature on the solid ($S$) or liquid ($L$) side
of the interface; also, $L$ and $c_p$ are, respectively, the latent heat of freezing 
and the specific heat at constant pressure per unit volume, 
and $D$ is the thermal diffusivity (with both $c_p$ and $D$ taken here
to be the same in both phases).

Ivantsov looked for needle-shape steady-state solutions of 
Eqs. \ref{e1} and \ref{e2} growing at a constant velocity $V$.
The assumption that growth takes place in steady-state allows one to
replace $\partial T/\partial t$ in Eq. \ref{e1} by $-V\partial T/\partial z$
in a frame of reference moving with respect
to the sample at speed $V$,
where $+z$ is the growth direction. The constraint that the needle
shape is preserved in time implies that $V_n=V\cos \theta$, where
$\theta$ is the angle between $\hat n$ and the $z$-axis. Remarkably,
Ivantsov found that paraboloids of revolution are exact solutions
to this steady-state growth problem for an isothermal interface at
$T=T_M$. Moreover, he derived a relationship between
the P\'eclet number, $P=\rho V/(2D)$, where $\rho$ is the 
dendrite tip radius, and the dimensionless undercooling 
$\Delta T/(L/c_p)$.
Three decades later, 
Glicksman {\it et al.} \cite{Glietal}
demonstrated by accurate measurements 
in a transparent organic system
that this relation is well satisfied.  

The Ivantsov relation does not predict $V$
and $\rho$ independently, but only their product.
Consequently, theoretical studies subsequent to Ivantsov's work 
concentrated on
understanding how this velocity is uniquely determined by including
capillary and kinetic effects. These effects make  
the interface non-isothermal, with a temperature 
determined by the classic
Gibbs-Thomson condition augmented to allow for a departure
from local equilibrium
\begin{equation}
T_I=T_M-\frac{T_M}{L}\sum_{i=1,2}
\left[\gamma(\hat n)+\frac{\partial^2 \gamma(\hat n)}{\partial \theta_i^2}
\right]\frac{1}{R_i}-\frac{V_n}{\mu(\hat n)}.\label{e3}
\end{equation}
The second term on the right-hand-side of Eq. \ref{e3} 
represents the purely equilibrium  
effect of capillary forces 
on the temperature of a curved interface, where $\gamma(\hat n)$ is
the excess free-energy of the solid-liquid interface,
$\theta_i$ are the local angles between
the normal direction $\hat n$ and the two local principal
directions on the interface, and $R_i$ are the
principal radii of curvature. This term has the crucial effect
that it makes the interface stable against short-wavelength 
perturbations. The last term on the right-hand-side of Eq. \ref{e3} 
is the non-equilibrium undercooling of the interface that 
drives crystallization by attachment  
of atoms at the interface, where $\mu(\hat n)$ is a kinetic coefficient. 
The solution of the steady-state growth problem  
becomes highly non-trivial when the isothermal condition
$T=T_M$ is replaced by Eq. \ref{e3}, and
several dendrite growth theories
after Ivantsov (reviewed in
Ref. \cite{TriKar}) made some uncontrolled approximations
that were later proven to be incorrect.

The fundamentally correct solution to
this problem emerged in the early nineteen eighties
from the study of simplified models where the interface dynamics is 
governed by local growth laws (a purely geometrical model 
where $V_n$ depends on the curvature and its second derivative
with respect to arclength along the interface \cite{Kesetal:geom},
and a boundary-layer model \cite{Benetal:blm}
that is an approximation of the dendritic growth equations in the limit
where the thickness of the thermal diffusion layer surrounding the tip 
is small compared to $\rho$). These models had the advantage that they
could be simulated numerically and studied analytically without uncontrolled
approximations that had been made in previous theoretical attempts to tackle directly
the dendrite growth equations with the full non-locality of
the diffusion field. Building on insights gained
from these studies, solutions to the
the full equations were then subsequently obtained
in two \cite{Lan:sol,KesLev:2D,Mei:2D,Ben:2D,BarLan}
and three dimensions \cite{KesLev,BenBre}.

The key insight gained from
these studies is that dendritic evolution
is controlled by a delicate balance between microscopic  
capillary effects and macroscopic heat transport in which
the anisotropic properties of the solid-liquid interface
play a crucial, and mathematically subtle, role. 
First, capillary forces act as a singular 
perturbation that transforms the Ivantsov continuous 
family of needle crystal solutions into a discrete set
of solutions. Second, capillary or/and kinetic 
effects must be anisotropic for solutions to exist; in this case
only the solution with the largest velocity in this discrete set
is linearly stable \cite{KesLev:stab} and therefore a potential 
candidate for the observed dendrite tip.  

The self-consistent calculation used to determine this velocity involves
a solvability condition, which is simply the condition
that a physically admissible solution to the steady-state growth equations
must be smooth at the tip. Consequently, the term solvability theory
is commonly used to refer to this type of calculation whether it is  
carried out numerically using a boundary integral approach
\cite{KesLev:2D,Mei:2D,Ben:2D,KesLev}
or analytically \cite{Lan:sol,BarLan,BenBre} 
using a WKB (Wentzel-Kramers-Brillouin) 
approximation in the limit of weak anisotropy.  

Over the last decade, new powerful computational approaches
have been developed to simulate the full time-dependent 
evolution of the solid-liquid interface governed by
Eqs. \ref{e1}-\ref{e3}. In particular, the 
phase-field method \cite{pfgen} has emerged as a powerful 
algorithm to simulate this evolution in both two and three dimensions
\cite{Kob,WarBoe,WanSek,KarRap,Proetal,Kimetal,KarRap:noise,PlaKar}.   
This approach has the well-known advantage that it facilitates front tracking
by making the phase boundaries spatially diffuse with the help of
scalar fields that naturally distinguish between different phases.
Using this approach, it has been possible 
to simulate the morphological development of dendrites and to  
validate quantitatively that the dynamically selected
dendrite tip operating state is indeed the one 
predicted by solvability theory \cite{KarRap}.

Despite this progress, our current ability to model
dendritic growth quantitatively for a given material remains very limited.
This is mainly due to the lack of knowledge  
of capillary and kinetic properties ($\gamma(\hat n)$ and $\mu(\hat n)$, 
respectively, which enter through Eq. \ref{e3}) and,
in particular, of their anisotropies 
that strongly influence the interface dynamics. Exceptions are 
some transparent organic materials \cite{HuaGli,GliSin,Musetal}
and only one metal system (Al-Cu) \cite{Tri:ani} for which 
values of the capillary anisotropy have been estimated from
experimental measurements of equilibrium shapes. 
Even for these materials, however, 
interface kinetic effects have not been quantified even though
they could be important. Traditionally, kinetic effects have been
assumed to be negligibly small compared to capillary effects
at small undercoolings, where the growth rate is small,
and the opposite for large
undercoolings. However, to predict dendritic evolution over
the entire experimentally accessible undercooling range requires a 
precise knowledge of both capillary and kinetic properties.

The hope to model dendritic evolution quantitatively
in metal systems has been recently revived by progress 
in modeling interfacial properties
using large scale molecular dynamics (MD) simulations
\cite{Hoyetal1,Hoyetal2,Moretal,Rametal}.
The interface kinetic coefficient has been calculated
for Ni and Cu by crystallizing large slabs of liquid with
interatomic potentials derived from the 
embedded atom method \cite{Hoyetal1}. Values of this coefficient
were found to be in good quantitative agreement with a model
of Broughton {\it et al.} \cite{Broetal} used previously to interpret  
simulations with the Lennard-Jones potential,
and about half an order of magnitude smaller than the value predicted
by a model of Turnbull \cite{Tur} that has been used
in the metallurgical literature to model rapid dendritic solidification.
Furthermore, the interface kinetics was found to be
strongly anisotropic, with growth at the same interface
undercooling being faster along [100] than [110], 
and faster along [110] than [111]. A similar magnitude of anisotropy has
also been found in simulations with the
Lennard-Jones potential \cite{Huietal}.

The anisotropy of the interfacial energy is more difficult to compute 
for materials with rough interfaces because it is extremely weak,
the maximum variation of $\gamma(\hat n)$ over all orientations
($\hat n$) of the interface being typically of the order of one percent.
Very recently, however, it has been demonstrated that this anisotropy
can be accurately computed by monitoring interfacial
fluctuations during MD simulations and
extracting the interfacial stiffness, i.e. $\gamma$ plus its
second derivative with respect to orientation in a given plane,
which is one order of magnitude more anisotropic 
than $\gamma$ itself \cite{Hoyetal2,Moretal}.

The goal of the present work is to model
dendritic growth in undercooled metallic melts fully quantitatively
by linking atomistic and phase-field simulations. 
We modify the existing phase-field formulation of
the crystallization of a pure melt in order to achieve an efficient
and accurate modeling of the large-undercooling regime
where interface kinetic effects become important,
and we use a recently developed
multi-scale algorithm \cite{PlaKar} to simulate this model in three dimensions.
Furthermore, we use as input into this model 
the anisotropic capillary and kinetic properties 
of the solid-liquid interface that have been recently
predicted by MD simulations \cite{Hoyetal1,Hoyetal2}. This allows us to make
quantitative predictions of morphological development that can be 
directly compared with experiments.

We focus on pure Ni for which the
interface properties have been most accurately modeled
to date by MD simulation. This system also has the advantage
of having been extensively studied experimentally. Various
groups have measured the relationship between the growth
velocity of the interface and the bulk undercooling
\cite{Wal,Wiletal,Hofetal,Lumetal,Mat}. Moreover, the morphological
evolution of the envelope of the solidification front during
recalescence (rapid freezing following triggered nucleation)
has been investigated by thermal imaging \cite{Lumetal,Mat}. 

We write down the phase-field model and examine
its sharp interface limit in section II. We then 
discuss the numerical implementation of 
this model in section III. Next, we
discuss the results of phase-field simulations in section IV.  
We mainly focus here on the comparison of these results
with experimental observations. The comparison with 
the predictions of solvability theory is only briefly summarized and
will be presented in more detail elsewhere. 
Finally, conclusions and prospects 
are given in section V.
 
\section{Phase-Field Model}

\subsection{Basic equations}

Over the last decade, the phase-field method \cite{pfgen} has
emerged as a powerful computational approach to simulate 
morphological development during solidification
\cite{Kob,WarBoe,WanSek,KarRap,Proetal,Kimetal,KarRap:noise,PlaKar}.  
This method has the well-known advantage
that it avoids to explicitly track a sharp phase boundary
by making the interface region between solid and liquid 
spatially diffuse over some finite thickness.
The general form of the phase-field model that has
been widely used to simulate the solidification of a pure melt
is defined by the equations
\begin{eqnarray}
\tau(\hat n)\,\frac{\partial\phi}{\partial t}&=&
-\frac{\delta F}{\delta \phi}, \label{p1}\\
\frac{\partial u}{\partial t}
&=&D\,\nabla^2u+\frac{1}{2}\frac{\partial
p(\phi)}{\partial t},\label{p2}
\end{eqnarray}
where $\delta F/\delta \phi$ denotes the functional derivative with
respect to $\phi$ of the free energy of the two-phase system. The latter 
assumes the phenomenological form
\begin{equation}
F=\int d\vec x\,\left[\frac{W^2(\hat n)}{2}|\nabla\phi|^2+f_{dw}(\phi,u)\right],
\label{Fdef}
\end{equation}
where the integral in Eq. \ref{Fdef} is over the volume of the
system and we have defined the dimensionless temperature field
\begin{equation}
u\equiv (T-T_M)/(L/c_p).\label{udef}
\end{equation}
Here, $f_{dw}$ is dimensionless and $W$ has dimension of length;
consequently, $F$ has dimension of volume.
It is important to emphasize that there is a significant
amount of freedom in the choice of the functional
form of the free energy density $f_{dw}(\phi,u)$
in the phase-field model. The first main requirement is  
that $f_{dw}$ should have the form of a double well potential with two minima 
corresponding to the liquid and solid phases, which are chosen
here at $\phi=\pm 1$ with the plus (minus) sign corresponding
to the solid (liquid). Thermodynamically, this guarantees
the existence of a well-defined
solid-liquid interface of thickness $\sim W$, i.e. where $\phi$
varies between $+1$ and $-1$ over this
thickness along a direction that is locally normal
to the interface. The second physical
requirement is that the difference in bulk
free-energy between solid and liquid, 
which drives the phase transformation, 
should be a monotonously increasing function
of the interface undercooling; 
this is equivalent, here, to requiring that
$f_{dw}(+1,u)-f_{dw}(-1,u)$ be a monotonously increasing 
function of $u$. A third requirement is that
the two minima of $f_{dw}$ remain at $\pm 1$
for different temperatures above and below the
melting point. A choice that 
satisfies all three requirements 
is \cite{WanSek,KarRap,Proetal,Kimetal,KarRap:noise,PlaKar,CagChe}
\begin{equation}
f_{dw}(\phi,u)=f(\phi)+\lambda \, u \, g(\phi),\label{fdef1}
\end{equation} 
with $\lambda$ a dimensionless constant, 
$f(\phi)=-\phi^2/2+\phi^4/4$, and  
\begin{equation} 
g(\phi)=\phi-2\phi^3/3+\phi^5/5. \label{gdef}
\end{equation} 
This form has been widely used to model 
dendritic solidification quantitatively in a low
undercooling regime where interface kinetic effects are  
vanishingly small \cite{KarRap,Proetal,PlaKar}. Here, however, we focus on the 
opposite high undercooling regime where kinetic effects are
dominant. In order to model accurately the kinetics of
the interface in this regime, it turns out to be necessary to
use the more general form
\begin{equation}
f_{dw}(\phi,u)=f(\phi)+h(\lambda \, u) \, g(\phi),\label{fdef2}
\end{equation}
which still satisfies the above requirements
if $h$ is a monotonously increasing function of its argument.
Even on today's computers and with
parallel codes, phase-field computations are only feasible for
values of $W$ much larger than the real atomic-scale width 
of the solid-liquid interface, denoted here by $\delta$.
With the double well potential defined by Eq. \ref{fdef1} and
$W\gg \delta$, the velocity of a planar interface 
in the phase-field model is a nonlinearly increasing function
of the interface undercooling for the high
velocity range ($30-60$ m/sec) that has been measured experimentally 
in Ni \cite{Wal,Wiletal,Hofetal,Lumetal,Mat}.
In contrast, atomistic simulations of the same material \cite{Hoyetal1} 
have shown that the relationship between the growth rate and the interface
undercooling is still linear in this velocity range. 
This problem can be solved by introducing a function $h$
in the double well potential defined by Eq. \ref{fdef2}
that allows one to freely choose the relation
between the interface undercooling and the thermodynamic
driving force for the phase transformation. Hence, 
even when the relationship between
the interface velocity and the driving force is nonlinear,
one can model a linear dependence of the
velocity on the interface undercooling. 
We discuss next how to explicitly compute
$h$ and how to choose the parameters  
in the phase-field model
by examining the mapping of this model onto
the sharp-interface equations of the previous section. 
The incorporation of thermal fluctuations
into the model is discussed in section III together
with the numerical implementation.

\subsection{Relationship between phase-field and sharp-interface models}

To examine the relationship between the phase-field and the
sharp-interface equations (Eqs. \ref{e1}-\ref{e3}), 
it is useful to first rewrite the latter equations 
in terms of $u$, which yields
\begin{eqnarray}
\frac{\partial u}{\partial t}&=&D\nabla^2 u,\label{s1}\\
V_n&=&D\,\hat n\cdot \left(\left.\vec\nabla u\right|_S
-\left. \vec\nabla u\right|_L\right) \label{s2},\\
u_I&=&-d_0\sum_{i=1,2}
\left[a_c(\hat n)+\frac{\partial^2 a_c(\hat n)}{\partial \theta_i^2}
\right]\frac{1}{R_i}-\beta(\hat n)V_n\label{s3},
\end{eqnarray}
where we have defined the microscopic capillary length
$d_0\equiv \gamma_0T_Mc_p/L^2$, which is typically
on the order of one nanometer,
and the kinetic coefficient 
\begin{equation}
\beta(\hat n)\equiv \frac{c_p}{L\,\mu(\hat n)}.\label{betdef}
\end{equation}
Here, $\gamma_0$ is the mean value of the
interfacial energy that can be defined by writting
$\gamma(\hat n)\equiv \gamma_0 a_c(\hat n)$, where the small variation
of the interfacial energy with orientation
is contained in the dimensionless function $a_c(\hat n)$
defined in section III below. The interface energy of
the phase-field model (which has dimension of length
because $f_{dw}$ was chosen dimensionless),
\begin{equation}
\gamma_{pf}(\hat n) = W(\hat n) \int_{-1}^1 \sqrt{2[f(\phi)-f(\pm 1)]}\; d\phi,
\label{surten}
\end{equation}
depends only on the shape of the free energy function and
is proportional to $W$. Therefore, anisotropic
capillary effects are correctly reproduced in the phase-field 
model by letting the interface thickness  
have the same anisotropy as the interfacial energy \cite{Kob},
that is $W(\hat n)=W_0a_c(\hat n)$ where the direction normal to
the interface is defined by $\hat n=-\vec \nabla \phi/
|\vec \nabla \phi|$. 

The phase-field model is well-known 
to map onto the above set of sharp-interface 
equations in the limit where the interface 
thickness is small compared to the scale 
of the microstructure. The asymptotic analysis
necessary to construct this mapping is rather involved
(see Ref. \cite{KarRap}
and earlier references therein). We only mention here
the main results that are necessary for our computations 
and to determine $h$.
Let us start with the equilibrium Gibbs-Thomson condition.
A curved interface in
the phase-field model is stationary only when
the temperature $u$ is constant and the
driving force induced by the curvature is exactly
compensated by the free energy difference between
the two bulk phases, $f_{dw}(+1,u)-f_{dw}(-1,u)=\lambda u [g(+1)-g(-1)]$,
which yields
\begin{equation}
\lambda u [g(+1)-g(-1)]= - \sum_{i=1,2}
\left[\gamma_{pf}(\hat n)+\frac{\partial^2 \gamma_{pf}(\hat n)}{\partial \theta_i^2}
\right]\frac{1}{R_i}.\label{gttt}
\end{equation}
Using Eq. \ref{surten} and comparing the result to Eq. \ref{s3},
we deduce that
\begin{equation}
d_0=a_1\frac{W_0}{\lambda},\label{d0}
\end{equation}
where $a_1=\int_{-1}^1 \sqrt{2[f(\phi)-f(\pm 1)]}/[g(+1)-g(-1)]$
is a constant of order unity
($a_1=5\sqrt{2}/8$ for the forms of $f(\phi)$ 
and $g(\phi)$ used here).

Eq. \ref{d0} implies that for a given capillary length
$d_0$, which is the only fixed length scale in the
dendrite growth problem, 
one has the freedom to vary the interface thickness 
$W_0$ in the phase-field model since $\lambda$ is a free
parameter. This freedom is due to the fact that
the free-energy difference on the left-hand-side
of Eq. \ref{gttt} is proportional to $\lambda$ whereas
the surface energy on the right-hand-side is proportional
to $W_0$. The goal of choosing
$W_0$ as large as possible in order to make
the phase-field equations less stiff is equivalent 
to choosing $\lambda\sim W_0/d_0$ as large as possible.
How large $\lambda$ can be chosen, under the constraint
that the results be converged (i.e. independent of $\lambda$),
has been shown to depend on
the choice of assumption made in mapping the phase-field model onto the
sharp-interface equations \cite{KarRap}.  
In the classical asymptotic analysis of the 
\emph{sharp-interface} limit of the phase-field model,  
$\lambda$ is assumed to be vanishingly
small. This assumption is equivalent, physically, to assuming that the
temperature is constant in the diffuse interface region in the limit
where its thickness vanishes. The 
expression for the kinetic coefficient that results from this
analysis is 
\begin{equation}
\beta(\hat n)=a_1\frac{\tau(\hat n)}{\lambda W(\hat n)}~.\label{bet0}
\end{equation} 
 A distinct \emph{thin-interface} limit 
has also been analyzed where $\lambda$ is assumed 
to be of order unity \cite{KarRap}. This analysis computes
a correction to the interface kinetics that results
from the variation of temperature across the interface
width, which is assumed finite (or, in simple terms,
the interface is assumed to be ``thin" as opposed to sharp). This analysis
yields the modified expression for the kinetic coefficient \cite{KarRap}
\begin{equation}
\beta(\hat n)=a_1\frac{\tau(\hat n)}{\lambda W(\hat n)}
\left[1-a_2\lambda\frac{W^2(\hat n)}{D\,\tau(\hat n)}\right],\label{bet}
\end{equation} 
where $a_2$ is a constant of order unity that depends again on the functional
forms of the functions $f(\phi)$ and $g(\phi)$ ($a_2=0.6267$ for
the forms used here). This expression has the advantage that it 
remains valid for larger $\lambda$ than
the one derived in the classical sharp-interface limit.
Furthermore, it can even be used 
to make the interface kinetics vanish by choosing 
$\tau(\hat n)=\lambda a_2W^2(\hat n)/D$, which makes the square
bracket in Eq. \ref{bet} vanish.  

In the early stage of the present investigation, we used 
the standard form of the double well potential given
by Eq. \ref{fdef1} to
simulate dendritic evolution in two dimensions for the parameters of 
pure Ni discussed below, together with Eqs. \ref{d0}
and \ref{bet} to interpret the results. We investigated  
values of $\Delta T/(L/c_p)$ varying between 
0.2 and 0.6, which is the undercooling range of
interest to make contact with experiments.
We found that the dendrite tip velocity only became independent
of the interface thickness for $\lambda$ of order unity, 
which is equivalent to choosing this thickness 
comparable to the atomic-scale width $\delta$ of the real
solid-liquid interface. These simulations turned out to be extremely
costly and not feasible in three dimensions.
The results of the simulations were analyzed in order to elucidate
the cause of this poor convergence. Plots of the temperature profile
along the central axis of the dendrite revealed that the magnitude
of the temperature variation across the diffuse interface is small
compared to the magnitude of the total interface undercooling. The latter is
itself a significant fraction of the total bulk undercooling at this
large growth rate. Consequently,
the correction to the kinetic coefficient predicted by the thin-interface
limit is essentially negligible.
This is easily seen by rewriting Eq. \ref{bet} in the form
\begin{equation}
\beta =\beta_0 \left[1-
a_2\lambda \frac{d_0}{D\beta_0}\right],\label{betc}
\end{equation}
using Eq. \ref{d0}, where $\beta_0$ denotes the kinetic coefficient predicted
by the sharp-interface limit (Eq. \ref{bet0}), and where
we have neglected anisotropy for simplicity.
For the parameters of pure Ni predicted by the MD simulations \cite{Hoyetal1}, 
$d_0/(D\beta_0)\approx 1/90$, such that the second term inside
the square bracket in Eq. \ref{betc} is small compared to one for
$\lambda$ of order unity.

Further analysis of the results 
revealed that the linear relationship between interface undercooling
and velocity, $u_I=-\beta V_n$, 
breaks down for the velocity range of interest here.  
Since only the product $\lambda u$ enters the phase-field equation,
this breakdown occurs at smaller velocity as $\lambda$ 
is increased. This explains the dependence on $\lambda$ 
of our dendrite simulation results 
with $f_{dw}$ defined by Eq. \ref{fdef1}. 
To overcome this difficulty, we use the new form of
$f_{dw}$ defined by Eq. \ref{fdef2}. The velocity of 
a planar interface is uniquely related to the driving force, which
is $-\lambda u_I$ in Eq. \ref{fdef1}, but $-h(\lambda u_I)$ in Eq. \ref{fdef2}.
Hence, once we have determined the nonlinear relationship
between the interface velocity and
$h$, we can then choose $h(\lambda u_I)$ so as to recover a
linear relationship between interface velocity and $u_I$. Let us first
compute the nonlinear relationship between the interface
velocity and $h$. This can
be done by rewriting Eq. \ref{p1}, with 
$f_{dw}$ defined by Eq. \ref{fdef2}, in a frame translating along the
$+x$ direction at velocity $V_n$ and by making the change of
variable $y=x/W(\hat n)$, which yields
\begin{equation}
v\,\frac{d\phi}{dy}+\frac{d^2\phi}{dy^2}-f'(\phi)-h\,g'(\phi)=0,
\label{1d}
\end{equation}
where we have defined the dimensionless interface velocity
$v=V_n\tau(\hat n)/W(\hat n)$ and the primes on
$f$ and $g$ denote differentiation with respect to $\phi$. Physically admissible
solutions of Eq. \ref{1d} that exhibit a smooth and monotonous 
variation of $\phi$ across the diffuse interface
only exist for a unique $v$ for
a given value of $h$. They can be found
by first writing Eq. \ref{1d} as a set of two coupled first-order differential
equations, and then using a shooting method to find the value of $v$ that
satisfies the boundary conditions $\phi(\pm \infty)=\mp 1$. By repeating
this procedure for different values of $h$, one can compute
the function $v(h)$, which we plot in Fig. \ref{functions}.
To make the interface kinetics linear
for large $\lambda$, it now
suffices to choose the function $h(\lambda u_I)=v^{-1}(-\lambda u_I/a_1)$, 
where $v^{-1}$ is the inverse of the function $v$. This inverse function is
uniquely defined because $v$ is a monotonous function of its argument. Consequently, 
$v(h(\lambda u_I))=v(v^{-1}(-\lambda u_I/a_1))=-\lambda u_I/a_1$, such that finally
\begin{equation}
V_n=\frac{W(\hat n)}{\tau(\hat n)}\,v(h(\lambda u_I))=-
 u_I/\beta(\hat n),  \label{kin1}
\end{equation}
where $\beta(\hat n)$ is given by Eq. \ref{bet0}, as desired. For modeling
dendrites, the magnitude of $\lambda u_I$ only becomes large  
when the interface is undercooled $(u_I<0)$, whereas this magnitude
is small during melting $(u_I>0)$, e.g. the remelting 
of secondary branches. Therefore,
we choose $h(\lambda u_I)=v^{-1}(-\lambda u_I/a_1)$ for $u_I<0$ and
$h(\lambda u_I)=\lambda u_I$ for $u_I>0$. Finally, by 
extending the above analysis to a moving curved interface,
it can also be shown that the Gibbs-Thomson condition with
a linear kinetic correction term (Eq. \ref{s3}) 
is correctly reproduced with the new form of
$f_{dw}$ defined by Eq. \ref{fdef2}.

\section{Numerical implementation}

In summary, we simulate the phase-field model defined by Eqs. \ref{p1}
and \ref{p2} together with the double-well potential defined by Eq. \ref{fdef2},
$f(\phi)=-\phi^2/2+\phi^4/4$, $g(\phi)$ defined by Eq. \ref{gdef},
the function $h(\lambda u)$ defined by $h(\lambda u)=\lambda u$
for $\lambda u>0$ and constructed, for $\lambda u<0$, 
from the inverse of the function $v(\lambda u)$ computed by
solving Eq. \ref{1d} as outlined above; values of $h$ for $\lambda u<0$ are 
tabulated beforehand at equally spaced intervals and 
$h$ is computed by simple linear interpolation 
from these tabulated values during the simulations. The interfacial
energy and the kinetic coefficient are assumed to have the forms 
\begin{eqnarray}
\gamma(\hat n)/\gamma_0\equiv a_c(\hat n)&=&
(1-3\epsilon_c)\left[1+\frac{4\epsilon_c}{1-3\epsilon_c}
(n_x^4+n_y^4+n_z^4)\right],\label{ac}\\
\beta(\hat n)/\beta_0\equiv a_k(\hat n)&=&
(1+3\epsilon_k)\left[1-\frac{4\epsilon_k}{1+3\epsilon_k}
(n_x^4+n_y^4+n_z^4)\right]\label{ak},
\end{eqnarray}
consistent with the symmetry of a 
crystalline material with an underlying 
cubic symmetry, where $\epsilon_c$ ($\epsilon_k$) is
the magnitude of the capillary (kinetic) anisotropy.
With the above definitions, a positive value of
$\epsilon_c$ ($\epsilon_k$) favors growth
along the [100] directions, which requires the
interfacial energy (kinetic coefficient) to be maximal 
(minimal) along these directions.
Using the definition $W(\hat n)=W_0 a_c(\hat n)$ together with 
Eqs. \ref{d0}-\ref{bet0}, we obtain 
\begin{eqnarray}
W(\hat n)&=&  \frac{8\lambda \,d_0}{5\sqrt{2}}\,a_c(\hat n)\label{Wdef} \\
\tau(\hat n)&= & 
\frac{32 \lambda^2 \,d_0 \beta_0}{25}
\,a_c(\hat n)\,a_k(\hat n) \label{taudef}
\end{eqnarray} 
that are integrated in the phase-field model using 
$\hat n =-\vec \nabla \phi/|\vec \nabla \phi|$, or
\begin{equation}
n_x^4+n_y^4+n_z^4=
\frac{(\partial \phi/\partial x)^4+(\partial \phi/\partial y)^4
+(\partial \phi/\partial z)^4}{|\vec \nabla \phi|^4}
\end{equation}
in Eqs. \ref{ac}-\ref{ak}. 

Thermal fluctuations are incorporated
into the model by using the standard Langevin formalism that has been widely
used in studies of second order phase transitions \cite{Haletal}, and more
recently in phase-field models of 
solidification \cite{KarRap:noise,Eldetal,PavSek}. This formalism
consists in adding random variables to the right-hand side
of the phase-field equations, with the noise in Eq. \ref{p2}
written in such a way that the total heat is conserved. These
random variables are chosen to be uncorrelated in space
and time and to obey Gaussian distributions whose variances
are dictated by the fluctuation-dissipation theorem.
This formalism has been used previously by Karma and Rappel \cite{KarRap:noise}
to investigate the formation of sidebranches in dendritic solidification, 
and we follow here exactly the same procedure to include fluctuations.
The variances of the non-conserved and conserved
noises added to Eqs. \ref{p1} and \ref{p2}, respectively,
are determined by a single dimensionless 
experimental parameter 
\begin{equation}
F_{expt}=\frac{k_BT_M^2c_p}{L^2d_0^3},
\end{equation}
where $k_B$ is the Boltzmann constant, 
as detailed in Ref. \cite{KarRap:noise}.

Our simulations exploit a hybrid finite-difference 
diffusion Monte Carlo algorithm that integrates efficiently the 
phase-field equations \cite{PlaKar}. In a first domain $D_1$,
consisting of the solid plus a thick liquid layer
surrounding the solid-liquid interface, 
the stochastic phase-field equations (with the Langevin noise 
added to both equations as in \cite{KarRap:noise}) are time-stepped
using a standard explicit Euler method with a finite-difference 
representation of the spatial derivatives on a cubic lattice.
In a second domain $D_2$, consisting of the rest of the liquid, Eq. \ref{p2}
reduces to the heat diffusion equation that is solved using a large ensemble of 
random walkers. The walkers make progressively larger steps with increasing
distance away from the interface, and this adaptive stepping reduces the
computer time in a way that is similar to adaptive meshing
algorithms. The two methods are interfaced
at the conversion boundary between domains $D_1$ and $D_2$
in such a way that heat is exactly conserved.

All the details of this algorithm are described in Ref. \cite{PlaKar}
and need not be reproduced here. In addition,   
the finite-difference implementation of the phase-field equations in $D_1$,
which was used in Ref. \cite{PlaKar}, has
been described in detail in Ref. \cite{KarRap}.
Here we make two additional improvements to this implementation.  
First, we integrate the equation for $u$
(Eq. \ref{p2}) in $D_1$ on a twice coarser mesh than the 
equation for $\phi$ (Eq. \ref{p1}); values of $u$ on the fine mesh
are then simply calculated by interpolating between values of $u$
on the coarse mesh. Numerical tests show that this
approach gives essentially identical results 
as using a single mesh. Second, we write the evolution equation for $\phi$  
derived from Eq. \ref{p1} by inserting all the anisotropic factors
in the form 
\begin{eqnarray}
\,\tau({\hat n})\,\partial_t \phi&=& 
W_0^2\nabla^2\phi+
\,{\vec \nabla}\cdot\left(\left[W({\hat n})^2-W_0^2\right]
{\vec \nabla} \phi\right)
+\sum_{w=x,y,z}\frac{\partial}{\partial w} 
\left(|\vec \nabla \phi|^2 W({\hat n})
\frac{ \partial W({\hat n})}
{\partial [\partial \phi/\partial w]}\right)\nonumber\\
& & +\left(\phi-h(\lambda\,u)\, (1-\phi^2)\right)(1-\phi^2),
\label{p2b}
\end{eqnarray}
where the second and third terms on the right-hand side
above vanish in the isotropic limit $\epsilon_c\rightarrow 0$.
The spatial derivatives associated with these terms are
discretized using centered finite-difference expressions, as before \cite{KarRap}.
In contrast, the isotropic Laplacian (first term
on the right-hand-side above) is discretized using the 18-point formula
\begin{equation}
\left.\nabla^2 \phi\right|_{i,j,k}= \frac{1}{3\Delta x^2}
\left[\sum_{nn}\phi+\frac{1}{2}\sum_{nnn}\phi-12\phi_{i,j,k}\right],
\label{18point}
\end{equation}
where $\sum_{nn}\phi$ and  $\sum_{nnn}\phi$ denote 
the sums of $\phi$-values evaluated at the $6$ nearest-neighbor and the 
12 next-nearest-neighbor positions on the cubic lattice, respectively,
where $\Delta x$ is the lattice spacing (equal along $x$, $y$,
and $z$). The key property of this 18-point formula is that it makes the lattice
corrections to the lowest order forms of the capillary and kinetic
anisotropies defined by Eqs. \ref{ac} and \ref{ak} vanish at
order $\Delta x^2$.  
This can be seen by performing a Taylor expansion in
$\Delta x$ of the discretized Laplacian. For planar 
interfaces parallel to the (100), (110), and (111) 
planes this expansion reduces to the form
$\nabla^2\phi\approx \partial^2\phi/\partial n^2 
+A(\Delta x)^2\partial^4 \phi /\partial n^4+O(\Delta x^4)$,
where $n$ is the coordinate normal to the interface,
and $A$ is a numerical constant. For the 18-point formula,
$A$ is \emph{identical} for all three planes.
In contrast, $A$ has a different value in each plane if the Laplacian 
is discretized using the standard finite-difference 
expression that only involves the nearest-neighbor sum.
This, in turn, generates order $\Delta x^2$ lattice corrections to the capillary
and kinetic anisotropies that can be taken into account 
following the estimation procedure developed in Ref. \cite{KarRap}.
The use of Eq. \ref{18point} eliminates the need for this procedure to resolve
small physical anisotropies and pushes anisotropic effects due to
the lattice discretization to a higher order ($\Delta x^4$)  
where numerical tests show that they are small enough to be neglected.

Let us briefly comment on
the sources of noise in our present numerical algorithm. The noise produced by
the random walkers is only generated in $D_2$, whereas the Gaussian
thermal noise that is added through stochastic variables in 
the phase-field equations is only generated in $D_1$. 
It was shown in Ref. \cite{PlaKar} that the
walker noise has a negligible effect on the interface evolution
because its magnitude decreases by several orders of magnitude across the
liquid layer surrounding the interface. Most
importantly, this noise is insufficient to sustain sidebranching.
One issue, however, is whether adding stochastic forces
to the phase-field equations only in region $D_1$
(as opposed to the entire computation domain $D_1+D_2$, as would be the case if we
had used a finite-difference scheme everywhere) reduces the strength
of the thermal noise in a degree that affects the results. To investigate this
issue numerically, we repeated certain test simulations with different
values for the thickness of the liquid layer surrounding the interface, 
and found that the sidebranch amplitude remained approximately
the same provided that this thickness is chosen 
large enough (about 25-50 times the width of the diffuse
interface in the present study).
Therefore, we conclude that our results are not
significantly affected by only adding Langevin noise in $D_1$.

We use for material parameters of pure Ni:   
$T_M=1726$ K, $L=2.311 \times 10^9$ J/m$^3$, 
$c_p=5.313 \times 10^6$ J/(m$^3$K), $D=10^{-5}$ m$^2$/sec,  
$\gamma_0=0.326$ J/m$^2$, $\epsilon_c=0.018$, 
$\mu_{100}= 0.52$ m/sec, and $\mu_{110}= 0.40$ m/sec,
where the capillary and kinetic parameters
have been estimated by monitoring the fluctuations
of the solid-liquid interface during MD simulations 
\cite{Hoyetal1,Hoyt:private}. 
From these values, we obtain $L/c_p=435$ K, $d_0=5.56\times 10^{-10}$ m,
$\beta_0=c_p(1/\mu_{100}+1/\mu_{110})/(2L)=5.084 \times 10^{-3}$ sec/m, 
$\epsilon_k=(\mu_{100}-\mu_{110})/(\mu_{110}+\mu_{110})=0.13$,
and $F_{expt}=0.234$.

Simulations were carried out on a
$800 \times 520 \times 520$ lattice with the (100) growth direction
along the larger dimension. We fully exploited the cubic 
symmetry of the crystal to reduce 
the computation time as described in \cite{PlaKar}. Furthermore, we used
a spacing $\Delta x=0.8 W_0$ and a time step $\Delta t=0.004 \,\tau_0$,
where $W_0=8\lambda d_0/(5\sqrt{2})$ and $\tau_0=32\lambda^2d_0\beta_0/25$.
Note that the physical domain being simulated is actually the sum of
the two aforementioned domains $D_1$ and $D_2$, and $D_2$ extends
farther than the lattice because the walkers make unrestricted off-lattice walks. 
We used as initial condition a small spherical seed and
followed the morphological development of the interface up to the
point where $D_1$ is no longer contained inside the 3-d lattice. This
allowed sufficient time for the dendrite tip to reach a steady-state
velocity in the simulations for the parameters studied. Finally, we increased
$\lambda$ from 2 to 16 as $\Delta T/(L/c_p)$ was decreased
from $0.8$ to $0.2$, which yielded converged results that are reasonably 
independent of interface thickness (i.e. of $\lambda$).

\section{Results and discussion}

\subsection{Comparison with experiments}

We compare in Fig. \ref{pfexpt} the variation of the steady-state dendrite 
growth velocity with bulk undercooling obtained from the phase-field simulations
with two sets of experimental measurements \cite{Wiletal,Mat}. 
These measurements were made
on small high-purity Ni droplets (a few millimeters in diameter)
using a widely used experimental technique that consists of electromagnetically 
levitating the droplets in order to avoid container-wall-induced 
heterogeneous nucleation, which is essential 
to attain very large undercoolings.  

In these experiments, crystallization of the undercooled melt
is externally initiated by mechanical contact between the liquid droplet
and a trigger needle at a well-defined undercooling.
After nucleation, the solidification front propagates
outwards from the trigger point throughout the bulk of the droplet.
The velocity of this front is measured by recording the abrupt
change of surface temperature that is produced by its passage. 
Willnecker {\it et al.} \cite{Wiletal} measured the temperature
with photodiodes in two regions of the droplet and obtained
an average measure of the velocity from the time difference of
the arrival of the front.
Lum {\it et al.} \cite{Lumetal} and
Matson \cite{Mat} used a fast high-resolution video camera that images 
the thermal profile of the entire surface and that provides, in
addition to the velocity, information about the morphological 
evolution of the solidification front.

The comparison in Fig. \ref{pfexpt} shows that the phase-field
model predictions are in reasonably good quantitative agreement
with the measured values of the velocity up to a critical undercooling
$\Delta T_c\approx 200$ K. For $\Delta T>\Delta T_c$, the measured
solidification velocity increases more slowly with undercooling
than predicted by the model, especially for the data set
of Lum {\it et al.} \cite{Lumetal} and Matson \cite{Mat}. This break in the slope 
of the measured $V-\Delta T$ curve, which is absent
in the simulations, was first seen
by Walker \cite{Wal}, and its existence has since been confirmed
by numerous experiments \cite{Wiletal,Hofetal,Lumetal,Mat}.  

From the analysis of the thermal imaging data,
Lum {\it et al.} \cite{Lumetal} and 
Matson \cite{Mat} have concluded that this break is
correlated with a change of morphology of the  
envelope of the solidification front, from angular for
$\Delta T<\Delta T_c$, to spherical for $\Delta T>\Delta T_c$.
Note that the spatial resolution of these images ($64\times 64$ pixels
over an area of a few millimeter square) is far too coarse to
image directly the interface on the scale of the
dendrite tip, which is in the submicron range for these high
undercoolings. Therefore, these images track the evolution of 
the macroscopic envelope of the solidification front,
but not of the front itself.

At these high growth rates, the thickness of the thermal diffusion boundary
layer (which scales as $D/V$) is small enough for the tips of surviving
secondary branches  
to grow independently, and at the same rate as primary branches.
Moreover, since secondary 
branches grow at 90$^0$ from the trunks of primary
dendrite branches sufficiently behind the tip, one would
expect the envelope of the solidification
front associated with steady-state dendrite growth along
the [100] directions to be pyramidal, or
angular once projected onto the droplet 
surface being imaged, as observed in Ref. \cite{Mat}.
In contrast, the observation of a transition to a 
smooth spherical envelope suggests that steady-state dendrite growth along
a preferred crystallographic direction no longer exists
for $\Delta T> \Delta T_c$. 

Our phase-field simulations show no indication of the existence
of such a transition with the present parameters used for Ni. The growth
morphology remains dendritic, with a well-defined steady-state
tip velocity, even for the largest simulated undercooling that is  
larger than $\Delta T_c$ in the experiments. Examples of
growth morphologies are shown in Figs. \ref{3dsnapshots}(a)-(b). Much larger
size simulations, which are not feasible in three dimensions, 
would be necessary to describe the dynamics of
the envelope of the solidification front on the scale where it
has been imaged experimentally. However, we expect this envelope to be pyramidal
over the entire range of undercoolings investigated here in view of
the smooth $V-\Delta T$ curve and the persistence of a 
stable dendrite tip for $\Delta T>\Delta T_c$. Larger size simulations
conducted in two dimensions for the same parameters  
confirm this expectation.

\subsection{Sensitivity to anisotropy and kinetic effects}

In order to investigate the source of the disagreement between
simulations and experiments for $\Delta T> \Delta T_c$,
and also to gain insight into the
factors controlling the interface dynamics,
we carried out additional simulations by varying independently
the magnitude of the capillary and kinetic anisotropy.
Varying the magnitude of the capillary anisotropy ($\epsilon_c$) 
was found to have very little influence on the results.
This is shown in Fig. \ref{vary} where we compare the phase-field
simulation results
for the value $\epsilon_c=0.018$ estimated for Ni \cite{Hoyetal1} 
and for $\epsilon_c=0$, corresponding to a purely isotropic
interfacial energy. The dendrite velocities are almost identical over
the whole range of undercooling investigated.  

In contrast, we found that both the 
solidification rate and the growth morphology
depend sensitively on the magnitude 
of the kinetic anisotropy ($\epsilon_k$) that was varied for a fixed
value of the capillary anisotropy $\epsilon_c=0.018$. 
This is shown in Fig. \ref{epskV} where we plot the computed
dendrite velocity as a function of $\epsilon_k$ at a fixed
undercooling. The velocity is seen to decrease as $\epsilon_k$ 
is lowered. Moreover, steady-state growth ceases
to exist for $\epsilon_k$ less than a
critical value $\epsilon_k^*$ that varies from zero to 
about 0.02 for $\Delta T/(L/c_p)$ varying from 
about $0.2$ to $0.7$ as shown in Fig. \ref{epskDelta}. We note
that $\epsilon_k^*$ should also depend on 
the capillary anisotropy that does not influence the
interface dynamics when $\epsilon_k$ is large, as already mentioned, 
but that does compete with the kinetic anisotropy  
when $\epsilon_k$ is small (i.e. comparable to $\epsilon_k^*$).
In particular, increasing $\epsilon_c$ shifts to higher
undercoolings the region where steady-state dendrite growth 
ceases to exist in the $\epsilon_k$-$\Delta T$ plane 
of Fig. \ref{epskDelta}.

For $\epsilon_k<\epsilon_k^*$, 
interface growth is dominated by tip-splitting,
which produces new branches, and branch-termination
by overgrowth from neighboring branches.
Since the branches do not have any obvious preferred growth 
direction, the envelope of the interface on a 
scale larger than the average branch spacing
becomes progressively more spherical as one moves away from the boundary that
marks the lower limit of existence of dendrite growth in the 
$\epsilon_k$-$\Delta T$ plane of Fig. \ref{epskDelta}. This
is illustrated in Fig. \ref{2dspacetime} that shows space-time
plots of the interface for two extreme values of $\epsilon_k$ at fixed $\Delta T$.
An example of a 3-d spherical envelope morphology is
shown in Fig. \ref{3dsnapshots}(c). This structure is qualitatively
similar to the so-called dense-branching morphology characteristic
of diffusion-limited growth with tip-splitting \cite{Benetal:DBM}. 
This morphology has been observed in a wide range of systems, including  
two-dimensional viscous flow in a Hele-Shaw cell
and amorphous annealing \cite{Benetal:DBM}. Fig. \ref{vary} shows that
the average velocity of the envelope of this morphology
is much slower here than the tip velocity of a dendrite grown  
the same undercooling with a large kinetic anisotropy.

The above results show that the interface dynamics is strongly
influenced by kinetic effects. This is due to the
fact that, for the large growth rates investigated here,
the interface kinetic undercooling ($\Delta T_I=\mu_{100}V$) is
a significant fraction of the total bulk undercooling, whereas
the curvature undercooling associated with capillary forces can
be estimated to be one to two orders of magnitude smaller 
than $\Delta T_I$ from the computed values of the tip radius.
Of course, one would expect the opposite to be true for
low undercoolings, but we have not investigated this range
here for the parameters of Ni.

\subsection{Comparison with solvability theory and previous numerical studies}

The phase-field results are in good quantitative agreement with the
predictions of an analytical solvability theory developed 
by Lee {\it et al.} that includes both 
capillary and kinetic effects \cite{Lee:thesis,Leeetal}. This theory is a direct
extension of the linearized solvability theory
developed previously by Barbieri and Langer \cite{BarLan}, which included
only capillary effects. In the theory of Lee {\it et al.}, the steady-state
growth equations are linearized around a paraboloid of revolution
and solved using a WKB approximation.
The resulting solvability condition for the existence of a smooth
dendrite solution is expressed in the form of a vanishing integral 
condition where the integral is evaluated numerically.
Details of this theory will be presented elsewhere and 
we only summarize here the main results.

We compare in Fig. \ref{vary} the dendrite velocity-undercooling curve
predicted by this theory, for the same parameters of Ni as used 
in the phase-field model, to the simulation results. The agreement is
remarkably good given the fact that the tip shape significantly deviates
from a paraboloid of revolution in this growth regime dominated
by anisotropic kinetics. Furthermore, this theory predicts
that dendrite growth ceases to exist in a region of the $\epsilon_k-\Delta T$
plane that is below the solid curve in Fig. \ref{epskDelta} and that
is in reasonably good quantitative agreement with
the phase-field results. The breakdown of the dendrite solution occurs 
because the main steady-state dendrite branch  
merges with a lower-velocity unstable branch at a critical
undercooling. Hence, dendrites only
exist below this critical undercooling as shown in
the morphology diagram of Fig. \ref{epskDelta}.

It should be emphasized that the transition from
dendritic growth to dense-branching morphology with
increasing undercooling is fundamentally
different in the present kinetic-dominated
regime than in the opposite capillary-dominated
regime that has been previously investigated in both 2-d 
\cite{Breetal} (with earlier references therein)
and 3-d \cite{Abeetal}.
In the latter regime, the disappearance of 
dendrites has been shown to be associated with noise-induced 
tip splitting \cite{Breetal}, or with the appearance of other branches
of steady-state solutions with a split tip \cite{Breetal,Abeetal}. 
In 2-d, the building block of the dense-branching or seaweed
morphology is the so-called doublon \cite{Breetal} consisting
of two asymmetric fingers that grow cooperatively. In a capillary
dominated growth regime, 
doublons exist above a critical undercooling that
depends on the magnitude of the capillary anisotropy. Beyond that threshold,  
the main steady-state branch of dendrite solutions still exists, but 
doublons grow faster and outrun dendrites. 
In 3-d, for growth in a channel, a triplet 
structure consisting of three asymmetric fingers 
has been found in numerical simulations \cite{Abeetal}.
No detailed quantitative study has been carried out, but it
might be expected that, qualitatively, 
triplets play the same role in 3-d as doublons in 2-d.
In contrast, in the present kinetic-dominated regime, we have
seen no evidence of doublons (triplets) in our 2-d (3-d) phase-field
simulations. Consequently, the emergence of the  
dense-branching morphology is directly linked to the 
\emph{termination} of the main dendrite steady-state branch  
at some critical undercooling \cite{Leeetal}. 

To further check this conclusion, 
we repeated {\it without noise} the
phase-field simulations corresponding to
isotropic kinetics and high undercoolings
in the morphology diagram of Fig. \ref{epskDelta}. 
We found that tip splitting still occurred.
Hence, we conclude that noise is not required here to induce
splitting. It is probable that noise shifts  
the critical undercooling at which splitting occurs, but
we have not conducted an exhaustive (and numerically
intensive) study to determine the magnitude of this shift. The relatively good
agreement between solvability and phase-field results 
in Fig. \ref{epskDelta} suggests that this shift is small.

In a recent 3-d phase-field simulation study that focused on
a capillary-dominated regime without kinetics \cite{Karetal:Flemings}, 
we found that the dendrite tip was destroyed
by thermal noise above a critical undercooling of about half the hypercooling
limit that roughly coincides with $\Delta T_c$ in the pure Ni experiments.
Furthermore, we conjectured that this noise-induced tip splitting
behavior could potentially explain the universally 
observed break in the $V-\Delta T$ curve, \emph{if} it persists
in the presence of kinetic effects \cite{Karetal:Flemings}.

The present study clearly demonstrates
that noise-induced tip splitting is prevented by 
strongly anisotropic kinetics for
the values of $\beta_0$ and $\epsilon_k$ estimated in the 
MD simulations, which invalidates this hypothesis. It is possible that
the values of these parameters could vary somewhat if the MD simulations  
were to be repeated with more accurate interatomic potentials. However,
$\epsilon_k$ would have to be reduced by about one order of magnitude for
a transition from dendrites to a spherical growth morphology to
occur with increasing undercooling according 
to the results of Fig. \ref{epskDelta}.
Moreover, since the phase-field results do agree quantitatively
with experiments for $\Delta T<\Delta T_c$, $\epsilon_k$ would
have to decrease abruptly close to $\Delta T_c$ to explain the data,
which seems unlikely. 

Another possibility is that $\beta_0$ is
not correctly estimated by the atomistic simulations, and that the
model could match the data over the whole undercooling range  
for a smaller value of $\beta_0$ and a very weak kinetic anisotropy.
This also seems unlikely because $\beta_0$ has been
extracted from MD simulations with different methods and potentials, 
which have all yielded results consistent with the 
Broughton-Gilmer-Jackson (BGK) model \cite{Broetal}. What remains
more uncertain is the magnitude of the kinetic anisotropy.

Very recently, Mullis and Cochrane \cite{MulCoc} carried out
2-d phase-field simulations of the crystallization of 
a pure melt using an estimate of $\beta_0$ based 
on the Turnbull model \cite{Tur} that
is several times smaller than the value  
predicted by MD simulations or the BGK model \cite{Broetal}. 
They also used a very weak kinetic anisotropy induced by
the orientation dependence of the interface thickness in
the phase-field model. They found that, for these kinetic parameters,
dendrites are destroyed by noise and replaced by a seaweed
structure at large growth rate. Furthermore, they conjectured,
as in Ref. \cite{Karetal:Flemings}, that this 
noise-induced tip splitting could be the mechanism of
the envelope transition observed by Matson \cite{Mat}. Again,
the present study invalidates this hypothesis, at least for
pure melts, under the assumption
that the kinetic parameters predicted by 
the MD simulations are correct.

Finally, the disappearance of the dendrite branch was not seen
in a previous numerical solvability calculation by Ben Amar \cite{Ben}  
with isotropic kinetics ($\epsilon_k=0$). 
The reason is that a much smaller kinetic coefficient
was used ($\beta_0D/d_0\approx 15$ in Ref. \cite{Ben} 
instead of $\beta_0D/d_0\approx 90$ here), which
extends the steady-state dendrite branch to higher undercoolings.

\subsection{Effects not included in the model}

If we assume that the existing MD estimates of interface
kinetic parameters are correct, then the most interesting conclusion 
from the present study is that other physical effects neglected here  
influence the interface dynamics for $\Delta T>\Delta T_c$. Since
we have used a phase-field model that describes the crystallization
of a pure substance with only diffusive transport in the melt,
two possible candidates are fluid flow and the presence
of very dilute impurities in the melt, which are both unavoidably 
present in experiments.

An analytical study of flow  
has been carried out by Horvay \cite{Hor}, motivated by 
experimental observations by Walker \cite{Wal2} in pure Ni.
He concluded that the negative pressure generated by
the density difference between solid and liquid should suffice
to induce cavitation during the very initial stage of crystallization
after nucleation. Although this phenomenon may be unrelated to the observed
morphology transition, it raises the issue of
whether fluid flow could modify sufficiently the conditions at the interface
(during the subsequent stage where the entire melt crystallizes) 
to alter the MD estimates of kinetic parameters,
and consequently the phase-field predictions.

Although the experiments to which we have compared
our phase-field results used high purity
Ni (e.g. better than 99.99\% in Ref. 
\cite{Wiletal}), some small amount of contamination may play a role.
Recently, Cochrane {\it et al.} have investigated the
effect of the oxygen content on the velocity-undercooling
curve in Cu \cite{Cocetal}. They found that the typically observed
break in the slope of this curve was absent for an oxygen
content less than 200 ppm, but present for Cu doped 
with 600 ppm of O. Furthermore, there is significant  
variation in the $V-\Delta T$ curves reported from various experiments
in Ni (see Fig. 1 in Lum {\it et al.}), which could potentially
result from oxygen contamination.

Even very dilute amounts of impurities can strongly influence 
the interface dynamics because solutal diffusion in the liquid is about four
orders of magnitude slower than heat diffusion. In addition,
impurities are known to be trapped by the rapidly advancing
interface for the velocities corresponding
to the undercooling range studied here \cite{Azi}. This
solute trapping effect need not be isotropic. If it is sufficiently
anisotropic, it could potentially alter the interface
dynamics. To our knowledge, however, nothing is known either experimentally
or theoretically about this anisotropy in metallic alloys.

\section{Conclusions}

The crystallization of highly undercooled Ni melts was modeled
using a computationally efficient phase-field approach together with  
capillary and kinetic properties of the solid-liquid interface
that have been recently estimated by atomistic 
simulations \cite{Hoyetal1,Hoyetal2}. In particular, 
the value used here for the kinetic coefficient ($\beta_0$) is
much larger than the estimate of the Turbull model \cite{Tur} that
has been traditionally used to model dendrite growth at high undercooling. 
Consequently, the interface dynamics is much more strongly dominated  
by kinetics, and our results are drastically different from those obtained  
in previous studies based on the Turbull estimate
\cite{Lumetal,Mat,MulCoc,Ben} or in related studies that 
focused on a purely capillary-dominated regime 
without kinetics \cite{Breetal,Karetal:Flemings}. 

The dependence of dendrite growth  
velocity on undercooling predicted
by the phase-field simulations is  
in good quantitative agreement with experimental data
for $\Delta T<\Delta T_c$, where $\Delta T_c$  is the critical undercooling 
corresponding to a break in the slope of the measured
$V-\Delta T$ curve. For $\Delta T>\Delta T_c$, the
velocity increases more rapidly with undercooling in the simulations, 
which do not show this break, than in the experiments.
Furthermore, the simulations predict that dendrite growth
persists even for the largest undercoolings that have been 
studied experimentally. In contrast, the 
thermal imaging data of Lum {\it et al.} \cite{Lumetal}
and Matson \cite{Mat} show that the envelope of the solid-liquid interface  
becomes spherical for $\Delta T>\Delta T_c$,
as opposed to angular for $\Delta T<\Delta T_c$, which suggests
that dendrites cease to exist for $\Delta T>\Delta T_c$ in
apparent disagreement with the present results.

Phase-field simulations in which the anisotropy parameters 
are systematically varied show that the growth rate and morphology
depend sensitively on the kinetic anisotropy. In particular, they show
that dendrites cease to exist above a critical undercooling 
comparable to $\Delta T_c$ if the kinetic anisotropy is about one order
of magnitude smaller than presently estimated by the atomistic
simulations. For $\Delta T>\Delta T_c$, these simulations yield
a dense-branching morphology characteristic
of tip-splitting-dominated growth. Furthermore, they show 
that the doublon (triplet) in 2-d (3-d) does not appear to be 
the underlying building block of this
morphology in this strongly kinetic-dominated regime, in contrast to
the seaweed structure that has been simulated in 2-d without kinetics \cite{Breetal} 
or with the much smaller value of $\beta_0$ 
based on the Turnbull model (and noise) \cite{MulCoc}. 

The phase-field results are in good quantitative agreement with the
predictions of a linearized solvability theory that includes both
capillary and kinetic effects \cite{Leeetal}. This theory
predicts dendrite velocities that are in remarkably good
quantitative agreement with phase-field results for the  
large kinetic anisotropy value predicted by the MD simulations. 
Furthermore, it predicts that for a weak kinetic anisotropy, the 
steady-state dendrite branch terminates at some critical
undercooling by merging with a lower velocity unstable branch.
This provides a natural explanation for the disappearance of
dendrites in the phase-field simulations with a sufficiently
weak kinetic anisotropy, even without noise.

These results emphasize the need for a precise knowledge of both
the magnitude of the kinetic coefficient  ($\beta_0$) 
and its anisotropy ($\epsilon_k$) to
model reliably morphological evolution in a kinetic-dominated growth regime.
There is still a significant uncertainty in
the value of $\epsilon_k$ obtained 
by MD simulations \cite{Hoyetal1}. The present
simulations, however, do show that $\epsilon_k$
must be large, as predicted \cite{Hoyetal1},
to reasonably fit the dendrite velocity data for 
$\Delta T<\Delta T_c$. This is, of course, if one uses the value of $\beta_0$ 
predicted by both MD simulations and the BGK model \cite{Broetal},
which is much less uncertain than the anisotropy. 

We are led to conclude that, 
if the present anisotropy value is confirmed, 
the widely observed velocity break 
in the $V-\Delta T$ curve \cite{Wal,Wiletal,Hofetal,Lumetal,Mat}
is not reproduced by the standard model of the solidification
of a pure melt with purely diffusive transport.
This suggests that this break, 
and the associated envelope
transition \cite{Mat}, may be caused by flow
effects or dilute impurities.
The latter seems possible since even very
dilute impurities in Cu have been shown to influence
the dendrite velocity and grain structure \cite{Cocetal}. 
An extension of the present study to link atomistic and
phase-field simulations in alloys 
is currently in progress to test this hypothesis and its consequences for
grain refinement \cite{Kar:gr}.

\noindent {\bf Acknowledgments}

This research was supported by the US Department of
Energy through grant No DE-FG02-92ER45471 and additional funds provided
by the Computational Materials Science Network.
Computations were carried out at NERSC and NU-ASCC.
We wish to thank Jeff Hoyt and Mark Asta for valuable
discussions and input.

\pagebreak

\begin{figure}
\caption{Plots of the dimensionless interface velocity $v$  
of an isothermal planar interface versus driving force 
$-h$ in the phase-field model.}
\label{functions}
\end{figure}

\begin{figure}
\caption{Dendrite velocity versus undercooling from two
sets of experiments in Ni by Lum {\it et al.} \protect\cite{Lumetal}
(open squares) and Willnecker {\it et al.} \protect\cite{Wiletal} (open circles), 
and from the phase-field simulations (filled triangles and solid line)
with $\lambda=$16, 12, 8, 5, 4, 3, and 2 for
$\Delta T/(L/c_p)=$ 0.2, 0.3, 0.4, 0.5, 0.6, 0.7, and 0.8, respectively,
and with the anisotropy parameters $\epsilon_c=0.018$ and
$\epsilon_k=0.13$  
estimated by atomistic simulations for pure Ni
\protect\cite{Hoyetal1,Hoyetal2}.}
 \label{pfexpt}
\end{figure}

\begin{figure}
\caption{Snapshots of 3-d interfaces for different
anisotropy parameters: (a) $\lambda=30$, 
$\epsilon_k=0.13$, $\epsilon_c=0.018$,
and $\Delta T/(L/c_p)=0.2$; (b) $\lambda=20$, 
$\epsilon_k=0.13$, $\epsilon_c=0.018$,
and $\Delta T/(L/c_p)=0.7$; (c) $\lambda=20$,
$\epsilon_k=0$, $\epsilon_c=0.018$, and $\Delta T/(L/c_p)=0.6$.
Larger values of $\lambda$ were used in these simulations to obtain
more developed structures for illustrative purposes.}
\label{3dsnapshots}
\end{figure}

\begin{figure}
\caption{Dendrite velocity versus undercooling obtained
from the phase-field simulations for $\epsilon_k=0.13$ and $\epsilon_c=0.018$ 
(filled triangles and solid line), $\epsilon_k=0.13$
and an isotropic interfacial energy $\epsilon_c=0$ (open circles),
and average velocity of the dense-branching morphology
with $\epsilon_k=0$ and $\epsilon_c=0.018$ (filled circle).
The predictions of solvability theory (thick dashed line)
for $\epsilon_k=0.13$ and $\epsilon_c=0.018$ agree
quantitatively with the simulations.}
\label{vary}
\end{figure}

\begin{figure}
\caption{Dendrite tip velocity (filled triangles)
and average velocity of the envelope of the interface 
(filled circles) versus kinetic anisotropy for $\lambda=4$,
$\epsilon_c=0.018$, and
$\Delta T/(L/c_p)=0.7$.}
\label{epskV}
\end{figure}

\begin{figure}
\caption{Growth morphology diagram showing the region of existence
of dendrite (filled triangles) and dense-branching (filled circles)
morphologies in the plane of kinetic anisotropy and undercooling for
the phase-field simulations with $\epsilon_c=0.018$. For the same parameters,
solvability theory predicts that steady-state dendrite growth solutions 
only exist above the solid line, in reasonably 
good quantitative agreement
with the phase-field results.}
\label{epskDelta}
\end{figure}

\begin{figure}
\caption{Plots of interface evolution for
strongly anisotropic (a) and isotropic (b) kinetics. 
Interfaces in (a) and (b) were extracted from the same simulations
that produced the dendrite and dense-branching morphology
shown in Figs. \protect\ref{3dsnapshots}(b)
and \ref{3dsnapshots}(c), respectively. The interfaces represent
cuts of the solid-liquid interface in a plane 
perpendicular to the [010] direction and are shown
at equal time intervals. The frame width and height are
both 8 $\mu$m in (a) and (b).}
\label{2dspacetime}
\end{figure}

\newpage

\vspace*{-0.2cm}
\centerline{FIG.1: J.Bragard {\em et al.} }
\vspace*{3.2cm}
\hspace*{3.2cm}

                \begin{center}
                \protect\epsfxsize=12.truecm
                \epsffile{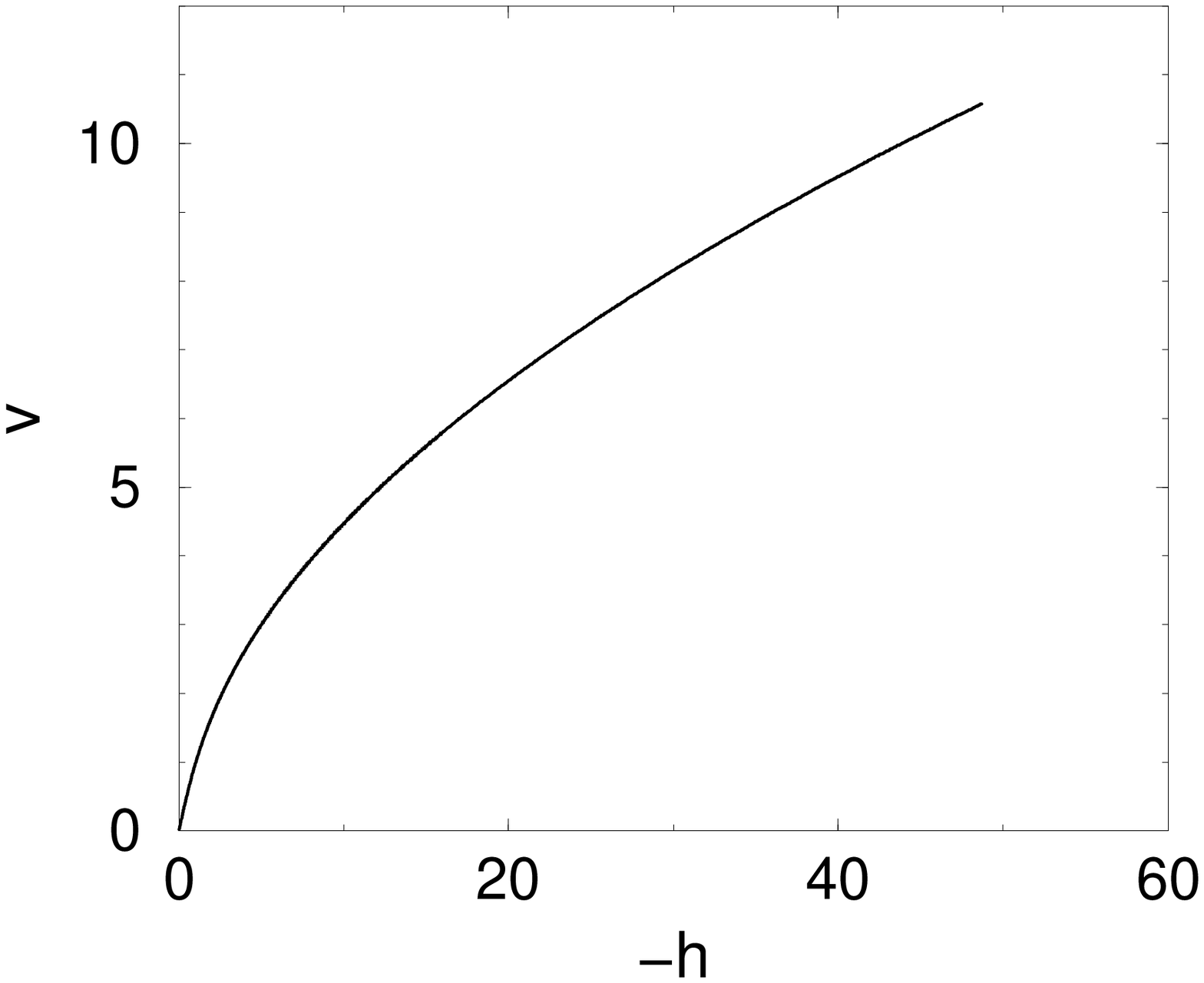}
                \end{center}

\newpage

\vspace*{-0.2cm}
\centerline{FIG.2: J.Bragard {\em et al.} }
\vspace*{3.2cm}
\hspace*{3.2cm}

                \begin{center}
                \protect\epsfxsize=12.truecm
                \epsffile{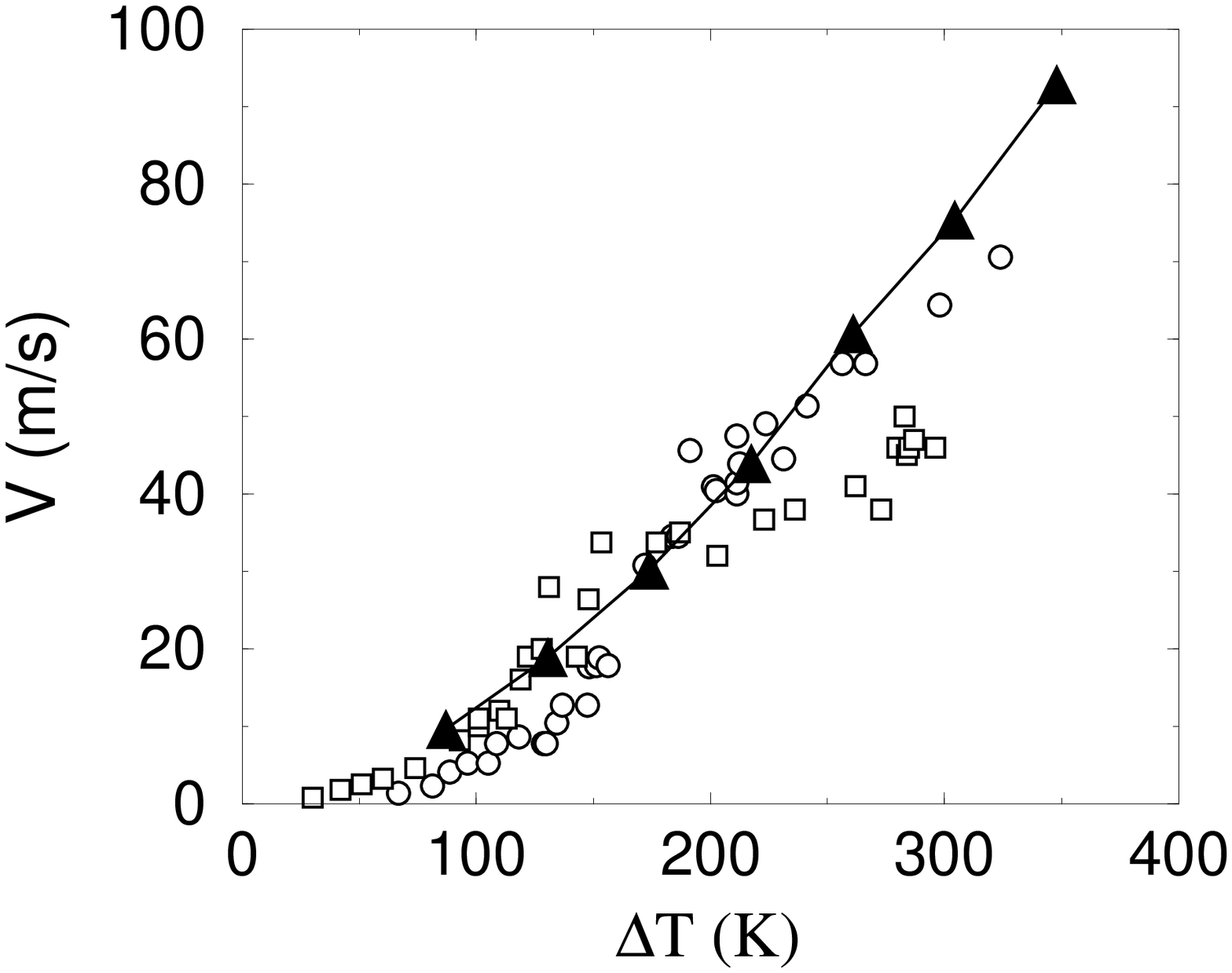}
                \end{center}

\newpage

\vspace*{-0.2cm}
\centerline{FIG.3a: J.Bragard {\em et al.} }
\vspace*{3.2cm}
\hspace*{3.2cm}

                \begin{center}
                \protect\epsfxsize=12.truecm
                \epsffile{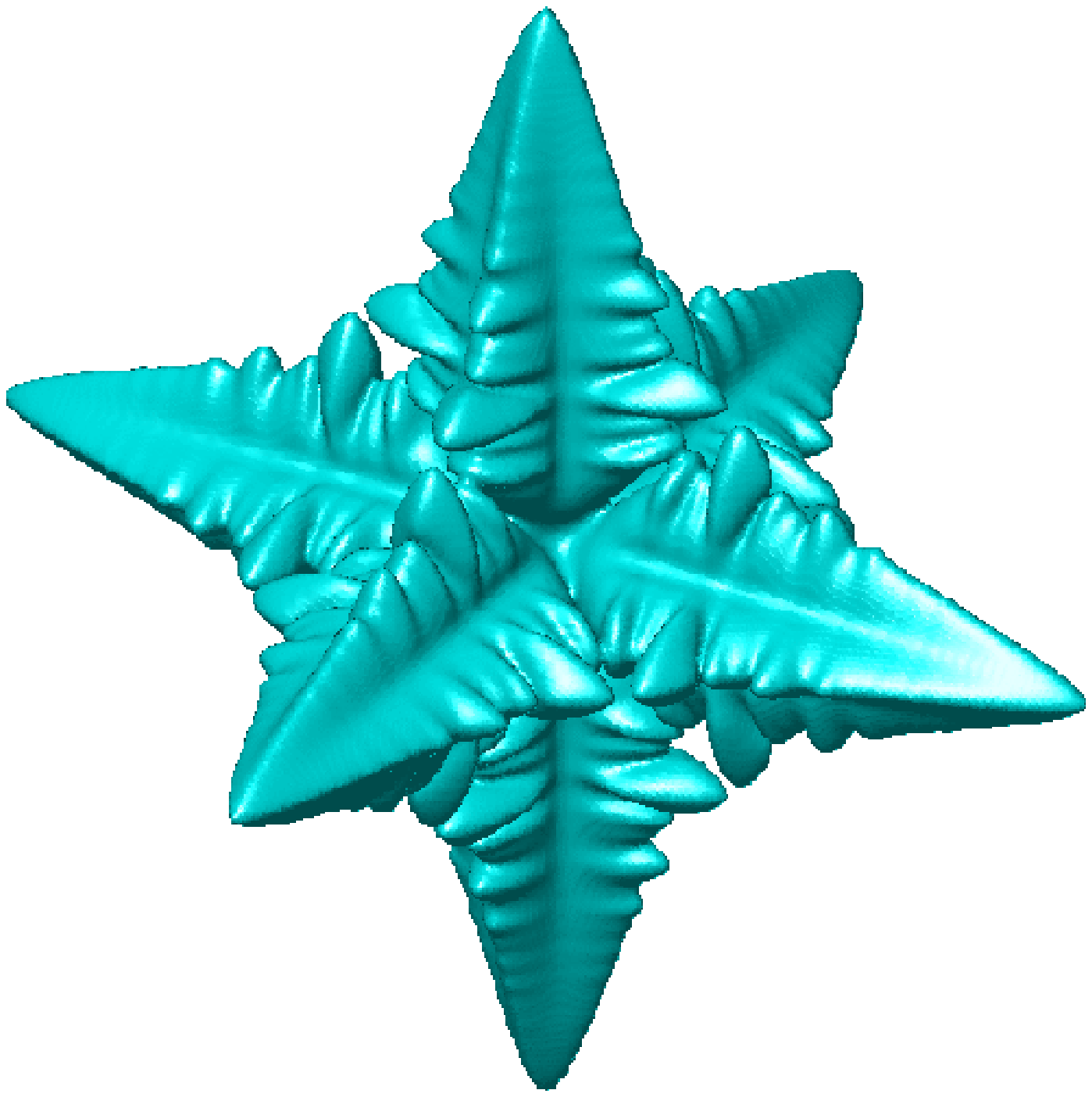}
                \end{center}

\newpage

\vspace*{-0.2cm}
\centerline{FIG.3b: J.Bragard {\em et al.} }
\vspace*{3.2cm}
\hspace*{3.2cm}

                \begin{center}
                \protect\epsfxsize=12.truecm
                \epsffile{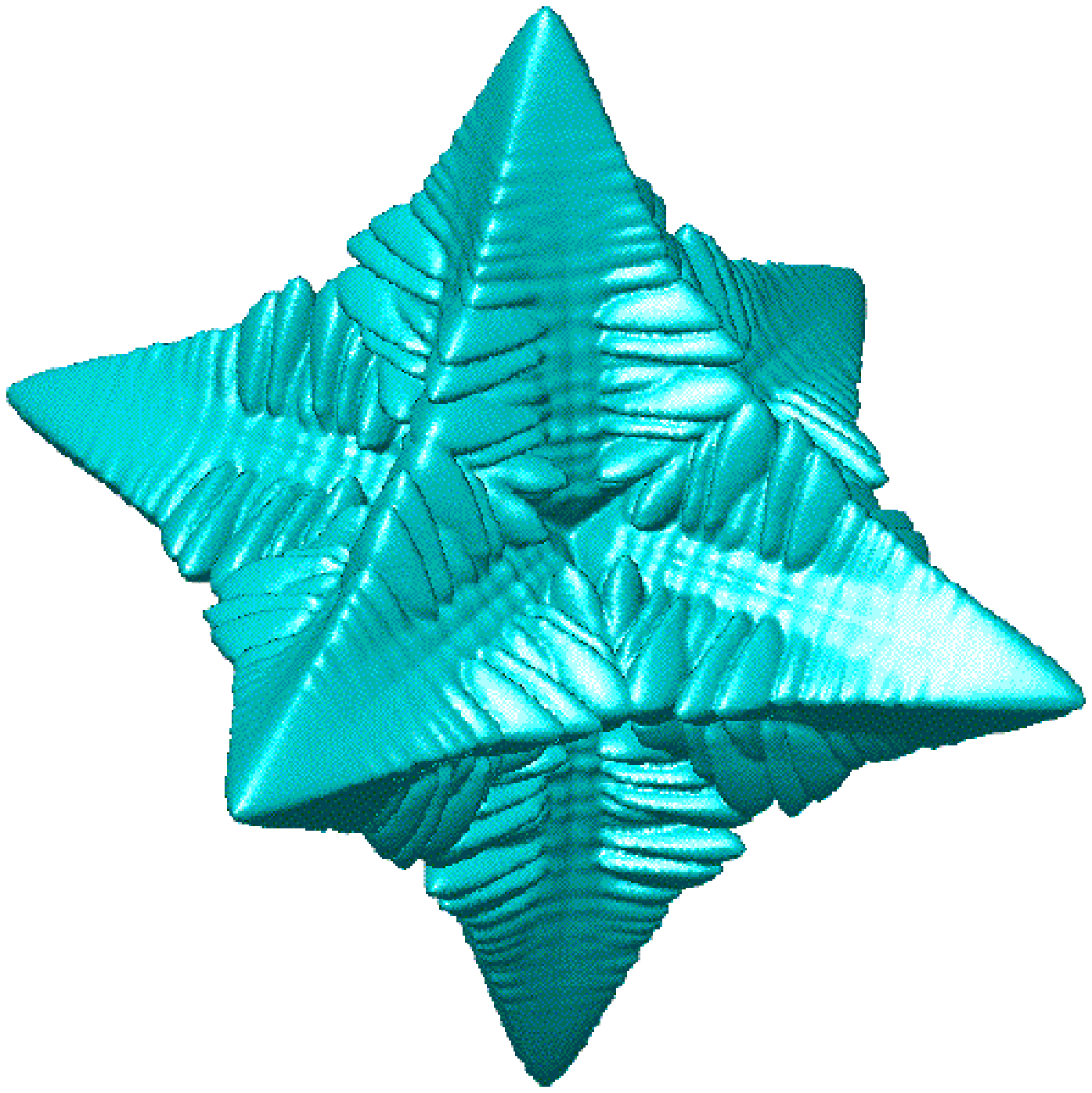}
                \end{center}

\newpage

\vspace*{-0.2cm}
\centerline{FIG.3c: J.Bragard {\em et al.} }
\vspace*{3.2cm}
\hspace*{3.2cm}

                \begin{center}
                \protect\epsfxsize=12.truecm
                \epsffile{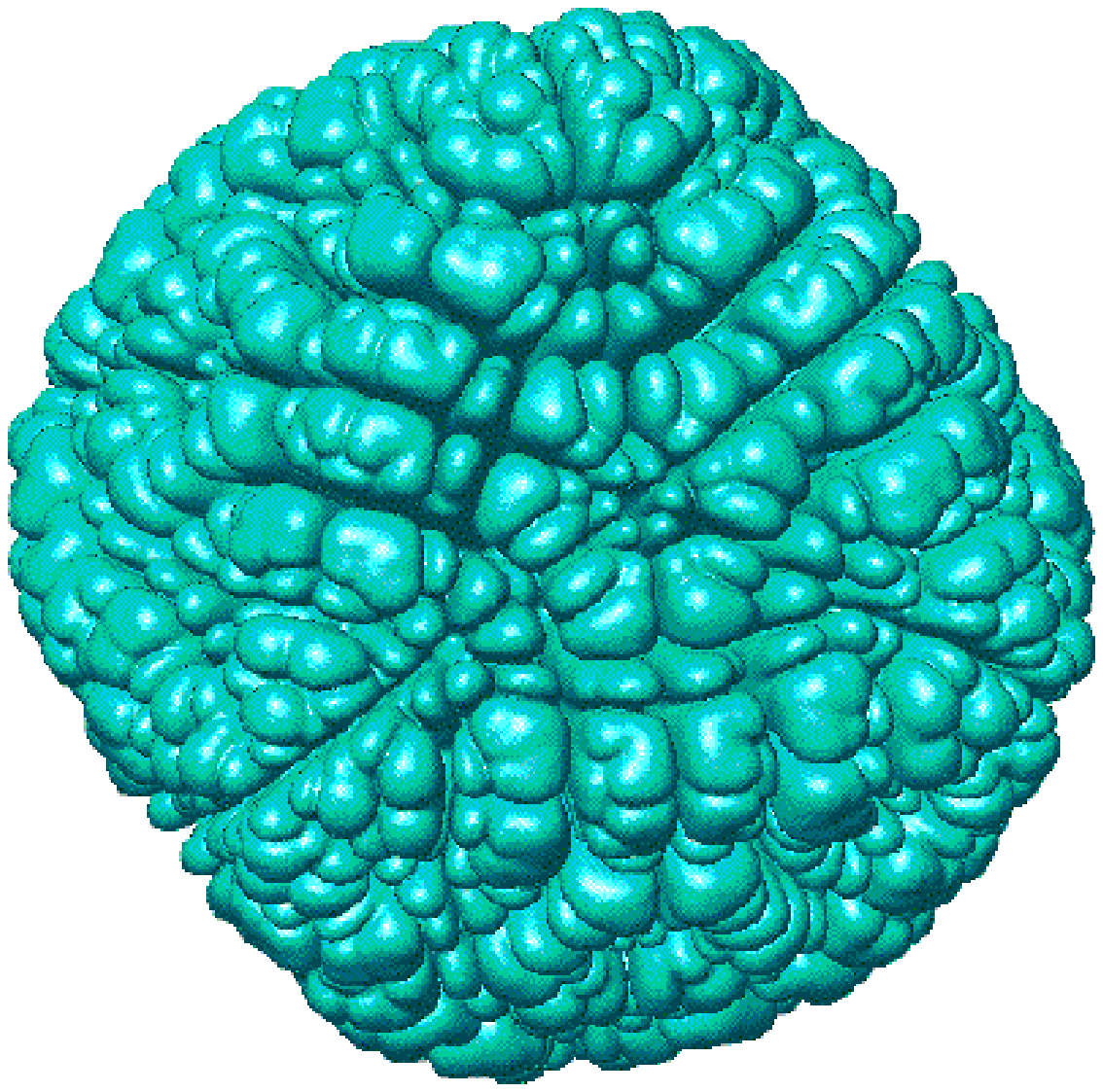}
                \end{center}

\newpage

\vspace*{-0.2cm}
\centerline{FIG.4: J.Bragard {\em et al.} }
\vspace*{3.2cm}
\hspace*{3.2cm}

                \begin{center}
                \protect\epsfxsize=12.truecm
                \epsffile{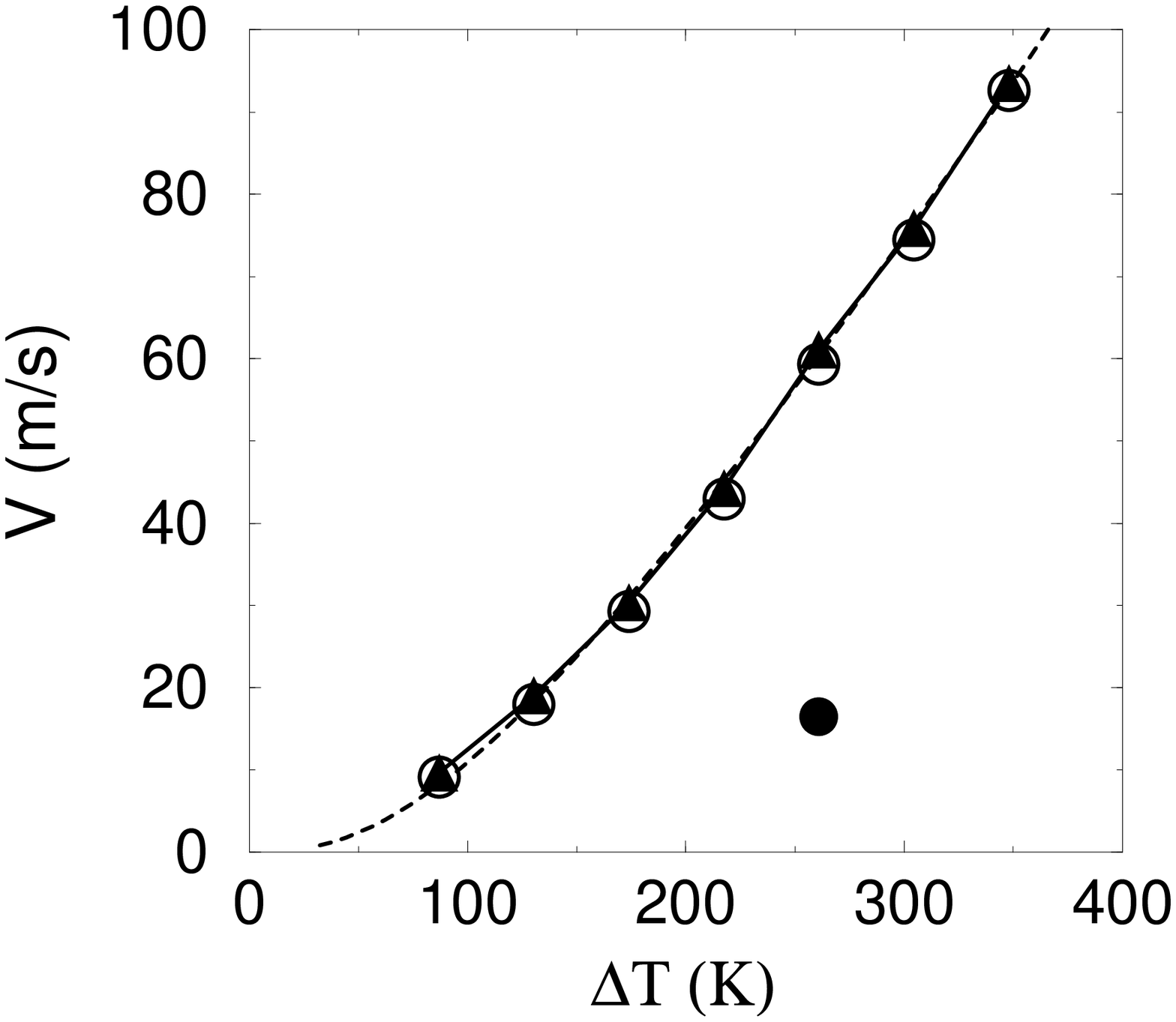}
                \end{center}

\newpage

\vspace*{-0.2cm}
\centerline{FIG.5: J.Bragard {\em et al.} }
\vspace*{3.2cm}
\hspace*{3.2cm}

                \begin{center}
                \protect\epsfxsize=12.truecm
                \epsffile{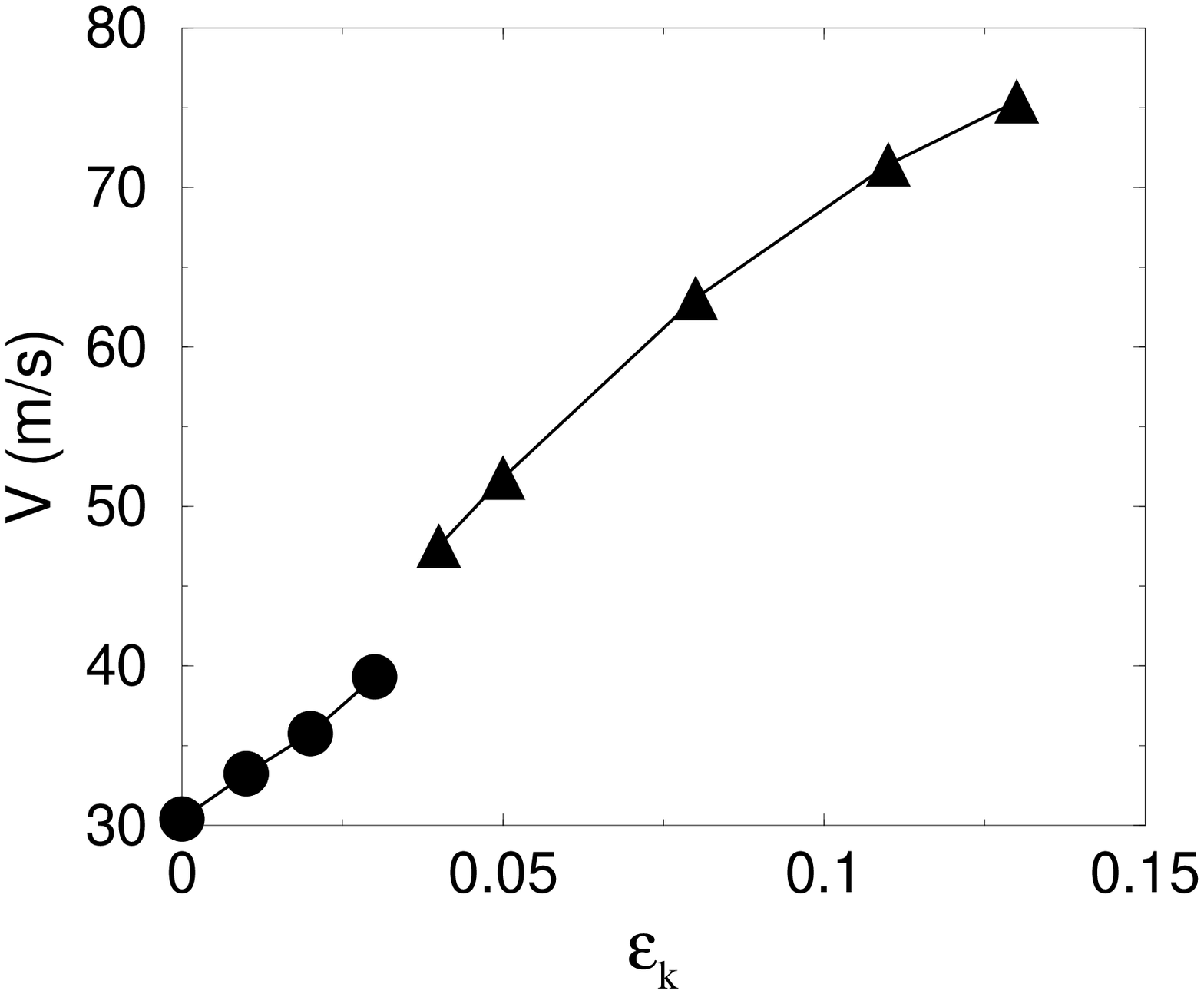}
                \end{center}

\newpage

\vspace*{-0.2cm}
\centerline{FIG.6: J.Bragard {\em et al.} }
\vspace*{3.2cm}
\hspace*{3.2cm}

                \begin{center}
                \protect\epsfxsize=12.truecm
                \epsffile{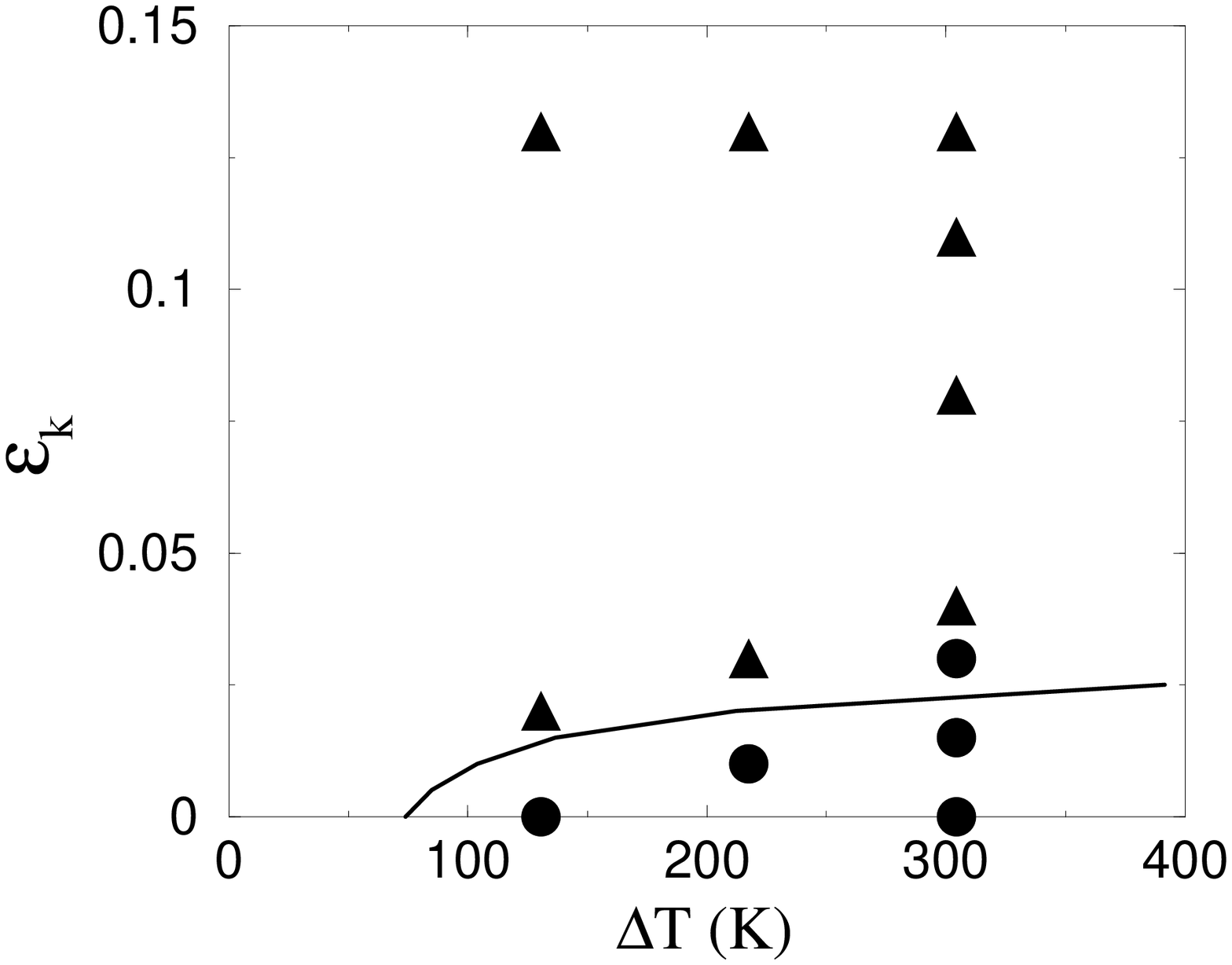}
                \end{center}

\newpage

\vspace*{-0.2cm}
\centerline{FIG.7a: J.Bragard {\em et al.} }
\vspace*{3.2cm}
\hspace*{3.2cm}

                \begin{center}
                \protect\epsfxsize=12.truecm
                \epsffile{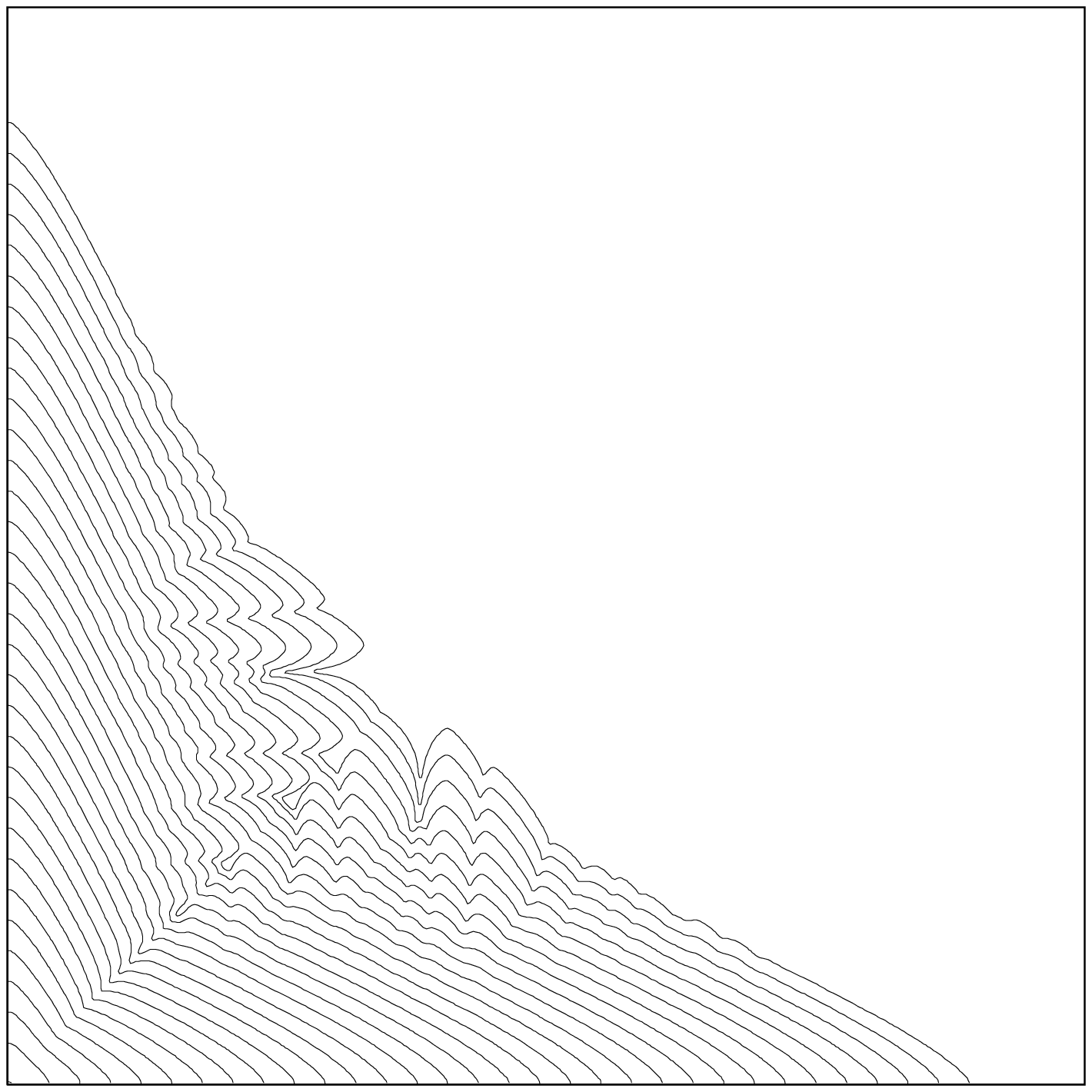}
                \end{center}

\newpage

\vspace*{-0.2cm}
\centerline{FIG.7b: J.Bragard {\em et al.} }
\vspace*{3.2cm}
\hspace*{3.2cm}

                \begin{center}
                \protect\epsfxsize=12.truecm
                \epsffile{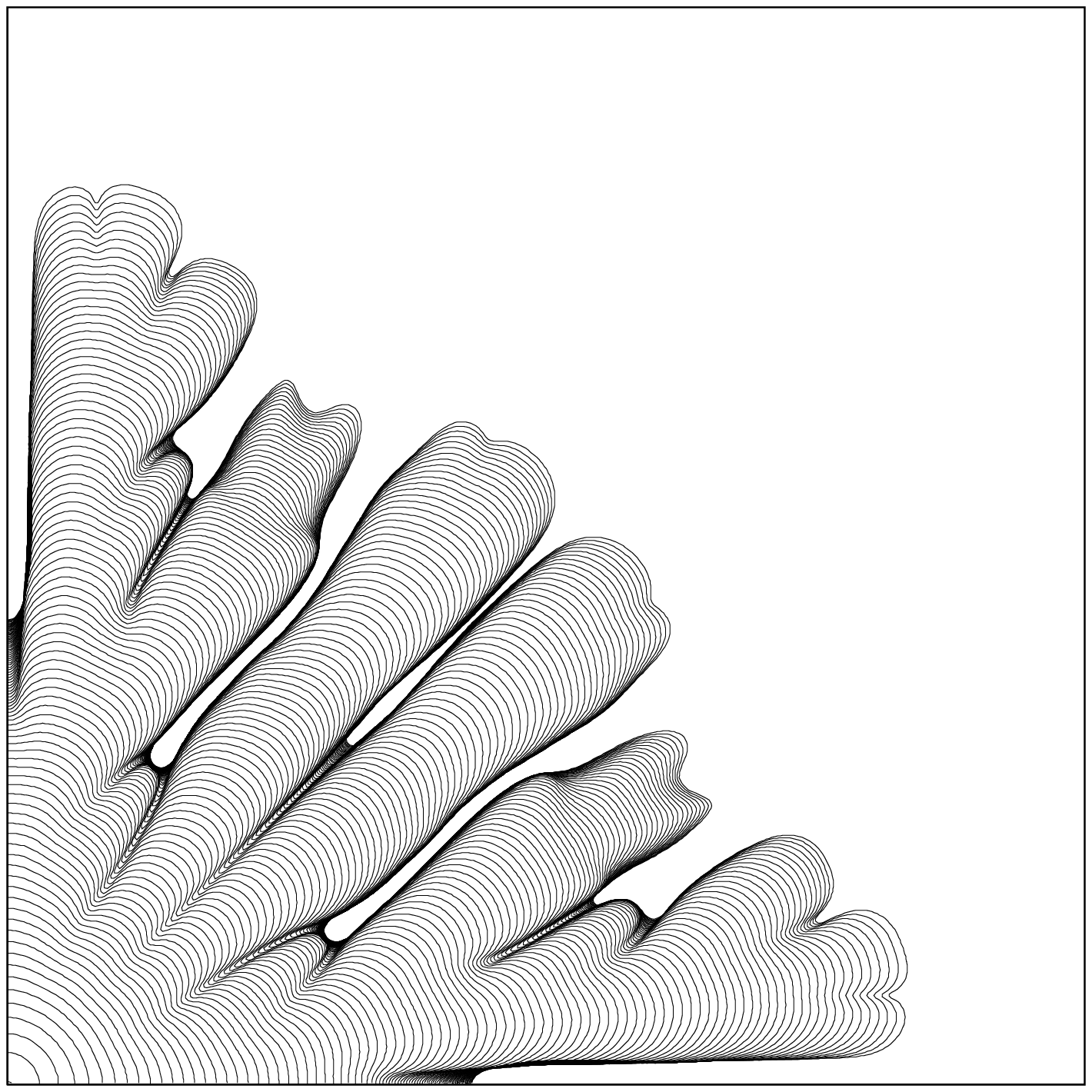}
                \end{center}

\end{document}